\documentclass[twocolumn,showpacs,preprintnumbers,amsmath,amssymb,superscriptaddress,prd]{revtex4}

\usepackage{graphicx}
\usepackage{epstopdf}
\usepackage{dcolumn}
\usepackage{bm}

\def \kprime{x}
\def \kprimeUncert{x}
\def \ExpectedSlope{x}
\def \CentralRegionSlope{x}

\def \deltahAfterStorage{x}

\def \limitNinety{x}
\def \limitNinetyfive{x}
\def \rangeNinetyLower{x}
\def \rangeNinetyUpper{x}
\def \rangeNinetyfiveLower{x}
\def \rangeNinetyfiveUpper{x}
\def \finalEdm{x}
\def \finalUncert{x}
\def \totalShiftExceptLight{x}
\def \totalShiftUncert{x}
\def \RxValue{x}
\def \RxValueBeforeRBSCorr{x}
\def \RxFitUncert{x}
\def \dxValue{x}
\def \dxUncert{x}
\def \dxUncertChisqCorr{x}
\def \RxFitUncertChisqCorr{x} 
\def \RxGlobalFit{x}
\def \RxGlobalFitUncert{x}

\def \RzeroUp{x}
\def \RzeroDown{x}
\def \Rxzero{x}  
\def \RxzeroUncert{x}
\def \dzero{x}
\def \dzeroUncert{x}
\def \dnfQEshift{x}		
\def \dnfQEcorrection{x}		
\def \dnfQEUncert{x}
\def \dfec{x}
\def \dfecUncert{x}
\def \ddip{x}
\def \Rdg{x}
\def \ddg{x}
\def \deltaDipShift{x}		
\def \deltaDipCorrection{x}	
\def \deltaDipUncert{x}
\def \RgammaFit{x}
\def \RgammaFitUncert{x}
\def \RqDown{x\pm x}
\def \RqUp{x\pm x}
\def \MagDipMom{x}
\def \MagDipDist{x}
\def \AlphaPeakShiftFromR0{x}
\def \AlphaPeakUp{x\pm x}
\def \AlphaPeakDown{x\pm x}
\def \DipFieldAtDoor{x}
\def \DipFieldAtFloor{x}
\def \dxChisqAdjustment{x}

\def \kprime{1.15} 			
\def \kprimeSoftMax{1.37} 	
\def \kprimeUncert{0.16} 		
\def \kprimeUncertSoftMax{0.19} 
\def \kprimeNoDubiousRun{0.93 \pm 0.19} 
\def \RxValueFromThreeParamLineFit{2.62 \pm 0.96} 
\def \dxValueFromThreeParamLineFit{-0.55 \pm 1.51} 
\def \ExpectedSlope{1.18}        
\def \CentralRegionSlope{1.18}	
\def \SlopesRatio{0.63} 		
\def \deltahAfterStorage{3.7}	
\def \limitNinety{3.0}			
\def \limitNinetyfive{3.6}		
\def \rangeNinetyLower{-3.2}	
\def \rangeNinetyUpper{2.8}	
\def \rangeNinetyfiveLower{-3.8}	
\def \rangeNinetyfiveUpper{3.4}	
\def \finalEdm{-0.21} 			
\def \finalUncert{1.82}			
\def \totalShiftExceptLight{-0.38}	
\def \totalShiftUncert{0.99}		
\def \RxValue{2.02}			
\def \RxValueBeforeRBSCorr{2.11}  
\def \RxFitUncert{0.89}		
\def \dxValue{-0.59}			
\def \dxUncert{1.53}			
\def \dxUncertChisqCorr{1.67}  	
\def \RxFitUncertChisqCorr{0.97} 
\def \RxGlobalFit{0.86}		
\def \RxGlobalFitUncert{1.08}	
\def \RzeroUp{1.14}			
\def \RzeroDown{0.58}		
\def \Rxzero{0.28}  			
\def \RxzeroUncert{0.19}		
\def \dnfQEshift{0.33}			
\def \dnfQEcorrection{-0.33}	
\def \dnfQEUncert{0.14}		
\def \dzero{-0.92}			
\def \dzeroUncert{1.68}		
\def \ddip{0.44}				
\def \Rdg{0.98}				
\def \ddg{1.15}				
\def \deltaDipShift{-0.71}		
\def \deltaDipCorrection{0.71}	
\def \deltaDipUncert{0.07}		
\def \dfec{-0.21}				
\def \dfecUncert{1.79}			
\def \RgammaFit{-0.22}		
\def \RgammaFitUncert{0.64}	
\def \RqDown{-0.5\pm1.1}		
\def \RqUp{2.7\pm1.1}		
\def \MagDipMom{-4.2}		
\def \MagDipDist{1.3}			
\def \AlphaPeakShiftFromR0{0.92}  
\def \AlphaPeakUp{0.2\pm1.1}	
\def \AlphaPeakDown{1.5\pm1.1}
\def \DipFieldAtDoor{1.1\ }		
\def \DipFieldAtFloor{0.3\ }		
\def \dxChisqAdjustment{0.68}	

\def \tabnumRvsH{I}  
\def \tabnumSystErrs{II}

\def \figabbr{Fig.\ }
\def \eqnabbr{Eq.\ }
\def \secabbr{Section\ }
\def \tababbr{Table\ }
\def \ecm{$e\,$cm}
\def \vecB{{\bf B}_0}

\def \dnf{d_\mathrm{n,Hg,f}}
\def \dBzdz{\partial B_z/\partial z}
\def \vxE{{\bf v}\times{\bf E}}
\def \a16{a_{16}}
\def \chinu{\chi^2/\nu}
\def \Rx{R_\times}

\def \dx{d_\times}

\begin{document}

\title{A Revised Experimental Upper Limit on the Electric Dipole Moment of the Neutron}

\author{J.M.~Pendlebury\footnote{Deceased}}
\affiliation{Department of Physics and Astronomy, University of Sussex, Falmer, Brighton BN1 9QH, UK}

\author{S.~Afach}
\affiliation{Paul Scherrer Institute, CH-5232 Villigen PSI,
Switzerland}
\affiliation{ETH Z\"{u}rich, Institute for Particle Physics, CH-8093 Z\"{u}rich, Switzerland}
\affiliation{Hans Berger Department of Neurology, Jena University Hospital, D-07747 Jena, Germany}

\author{N.J.~Ayres}
\affiliation{Department of Physics and Astronomy, University of Sussex, Falmer, Brighton BN1 9QH, UK}

\author{C.A. Baker}\affiliation{STFC Rutherford Appleton Laboratory, Harwell, Didcot, Oxon OX11 0QX, UK}

\author{G.~Ban}
\affiliation{LPC Caen, ENSICAEN, Universit\'{e} de Caen, CNRS/IN2P3, Caen, France}

\author{G.~Bison}
\affiliation{Paul Scherrer Institute, CH-5232 Villigen PSI, Switzerland}

\author{K.~Bodek}
\affiliation{Marian Smoluchowski Institute of Physics, Jagiellonian University, 30-348 Cracow, Poland}

\author{M.~Burghoff}
\affiliation{Physikalisch Technische Bundesanstalt, Berlin, Germany}






\author{P. Geltenbort}\affiliation{Institut Laue-Langevin, CS 20156 F-38042 Grenoble Cedex 9, France}

\author{K. Green}\affiliation{STFC Rutherford Appleton Laboratory, Harwell, Didcot, Oxon OX11 0QX, UK}

\author{W.C.~Griffith}
\affiliation{Department of Physics and Astronomy, University of Sussex, Falmer, Brighton BN1 9QH, UK}

\author{M. \surname{van der Grinten}}\affiliation{STFC Rutherford Appleton Laboratory, Harwell, Didcot, Oxon OX11 0QX, UK}

\author{Z.D.~Gruji\'c}
\affiliation{Physics Department, University of Fribourg, CH-1700 Fribourg, Switzerland}

\author{P.G.~Harris
\footnote{Corresponding author.  e-mail: p.g.harris@sussex.ac.uk}}\affiliation{Department of Physics and Astronomy, University of Sussex, Falmer, Brighton BN1 9QH, UK}


\author{V.~H\'{e}laine
\footnote{Present address: LPSC, Universit\'{e} Grenoble Alpes, CNRS/IN2P3, Grenoble, France}}
\affiliation{LPC Caen, ENSICAEN, Universit\'{e} de Caen, CNRS/IN2P3, Caen, France}

\author{P. Iaydjiev\footnote{On leave of absence from Institute of Nuclear Research and Nuclear Energy, Sofia, Bulgaria}}\affiliation{STFC Rutherford Appleton Laboratory, Harwell, Didcot, Oxon OX11 0QX, UK}

\author{S.N. Ivanov\footnote{On leave of absence from Petersburg Nuclear Physics Institute, Russia}}\affiliation{STFC Rutherford Appleton Laboratory, Harwell, Didcot, Oxon OX11 0QX, UK}

\author{M.~Kasprzak}
\affiliation{Physics Department, University of Fribourg, CH-1700 Fribourg, Switzerland}
\affiliation{Instituut voor Kern- en Stralingsfysica, Katholieke Universiteit Leuven, B-3001 Leuven, Belgium}

\author{Y.~Kermaidic}
\affiliation{LPSC, Universit\'{e} Grenoble Alpes, CNRS/IN2P3, Grenoble, France}

\author{K.~Kirch}
\affiliation{Paul Scherrer Institute, CH-5232 Villigen PSI, Switzerland}
\affiliation{ETH Z\"{u}rich, Institute for Particle Physics, CH-8093 Z\"{u}rich, Switzerland}


\author{H.-C.~Koch}
\affiliation{Physics Department, University of Fribourg, CH-1700 Fribourg, Switzerland}
\affiliation{Institut f\"{u}r Physik, Johannes-Gutenberg-Universit\"{a}t, D-55128 Mainz, Germany}

\author{S.~Komposch}
\affiliation{Paul Scherrer Institute, CH-5232 Villigen PSI, Switzerland}
\affiliation{ETH Z\"{u}rich, Institute for Particle Physics, CH-8093 Z\"{u}rich, Switzerland}

\author{A.~Kozela}
\affiliation{Henryk Niedwodnicza\'nski Institute for Nuclear Physics, Cracow, Poland}

\author{J.~Krempel}
\affiliation{ETH Z\"{u}rich, Institute for Particle Physics, CH-8093 Z\"{u}rich, Switzerland}
\affiliation{Paul Scherrer Institute, CH-5232 Villigen PSI, Switzerland}

\author{B.~Lauss}
\affiliation{Paul Scherrer Institute, CH-5232 Villigen PSI, Switzerland}

\author{T.~Lefort}
\author{Y.~Lemi\`{e}re}
\affiliation{LPC Caen, ENSICAEN, Universit\'{e} de Caen, CNRS/IN2P3, Caen, France}

\author{D.J.R.~May}
\affiliation{Department of Physics and Astronomy, University of Sussex, Falmer, Brighton BN1 9QH, UK}


\author{M.~Musgrave}
\affiliation{Department of Physics and Astronomy, University of Sussex, Falmer, Brighton BN1 9QH, UK}

\author{O.~Naviliat-Cuncic} \altaffiliation{Michigan State University, East-Lansing, USA.}
\affiliation{LPC Caen, ENSICAEN, Universit\'{e} de Caen, CNRS/IN2P3, Caen, France}

\author{F.M.~Piegsa}
\affiliation{ETH Z\"{u}rich, Institute for Particle Physics, CH-8093 Z\"{u}rich, Switzerland}

\author{G.~Pignol}
\affiliation{LPSC, Universit\'{e} Grenoble Alpes, CNRS/IN2P3, Grenoble, France}


\author{P.N.~Prashanth}
\affiliation{Instituut voor Kern- en Stralingsfysica, Katholieke Universiteit Leuven, B-3001 Leuven, Belgium}

\author{G.~Qu\'{e}m\'{e}ner}
\affiliation{LPC Caen, ENSICAEN, Universit\'{e} de Caen, CNRS/IN2P3, Caen, France}

\author{M.~Rawlik}
\affiliation{ETH Z\"{u}rich, Institute for Particle Physics, CH-8093 Z\"{u}rich, Switzerland}

\author{D.~Rebreyend}
\affiliation{LPSC, Universit\'{e} Grenoble Alpes, CNRS/IN2P3, Grenoble, France}

\author{J.D.~Richardson}\affiliation{Department of Physics and Astronomy, University of Sussex, Falmer, Brighton BN1 9QH, UK}

\author{D.~Ries}
\affiliation{Paul Scherrer Institute, CH-5232 Villigen PSI, Switzerland}
\affiliation{ETH Z\"{u}rich, Institute for Particle Physics, CH-8093 Z\"{u}rich, Switzerland}

\author{S.~Roccia}
\affiliation{CSNSM, Universit\'{e} Paris Sud, CNRS/IN2P3, Orsay, France}

\author{D.~Rozpedzik}
\affiliation{Marian Smoluchowski Institute of Physics, Jagiellonian University, 30-348 Cracow, Poland}

\author{A.~Schnabel}
\affiliation{Physikalisch Technische Bundesanstalt, Berlin, Germany}

\author{P.~Schmidt-Wellenburg}
 \affiliation{Paul Scherrer Institute, CH-5232 Villigen PSI, Switzerland}

\author{N.~Severijns}
\affiliation{Instituut voor Kern- en Stralingsfysica, Katholieke Universiteit Leuven, B-3001 Leuven, Belgium}

\author{D. Shiers}
\affiliation{Department of Physics and Astronomy, University of Sussex, Falmer, Brighton BN1 9QH, UK}

\author{J.A.~Thorne}
\affiliation{Department of Physics and Astronomy, University of Sussex, Falmer, Brighton BN1 9QH, UK}


\author{A.~Weis}
\affiliation{Physics Department, University of Fribourg, CH-1700 Fribourg, Switzerland}

\author{O.J.~Winston}
\affiliation{Department of Physics and Astronomy, University of Sussex, Falmer, Brighton BN1 9QH, UK}

\author{E.~Wursten}
\affiliation{Instituut voor Kern- en Stralingsfysica, Katholieke Universiteit Leuven, B-3001 Leuven, Belgium}


\author{J.~Zejma}
\affiliation{Marian Smoluchowski Institute of Physics, Jagiellonian University, 30-348 Cracow, Poland}


\author{G. Zsigmond}
\affiliation{Paul Scherrer Institute, CH-5232 Villigen PSI, Switzerland}

\date{\today}

\begin{abstract}

We present for the first time a detailed and comprehensive analysis of the experimental results that set the current world sensitivity limit on the magnitude of the electric dipole moment (EDM) of the neutron.  We have extended and enhanced our earlier analysis to include recent developments in the understanding of the effects of gravity in depolarizing ultracold neutrons (UCN); an improved calculation of the spectrum of the neutrons; and conservative estimates of other possible systematic errors, which are also shown to be consistent with more recent measurements undertaken with the apparatus.   We obtain a net result of $d_\mathrm{n} = \finalEdm \pm \finalUncert \times10^{-26}$ \ecm, which may be interpreted as a slightly revised upper limit on the magnitude of the EDM of $\limitNinety  \times10^{-26}$ \ecm\ (90\% CL) or $ \limitNinetyfive  \times10^{-26}$ \ecm\ (95\% CL).  

This paper is dedicated by the remaining authors to the memory of Prof. J.\ Michael Pendlebury.

\end{abstract}

\maketitle
\section{Introduction}
\label{sec:intro}

Measurement of the electric dipole moment (EDM) of the neutron (or of other fundamental particles) provides an extremely sensitive approach to investigating potential new sources of CP violation and physics beyond the Standard Model.

The experimental technique underlying the most sensitive measurement to date \cite{Baker_2006, Baker_2007} of the neutron EDM has been discussed extensively in an earlier publication\cite{Baker_2014}.  The data were collected at ILL, Grenoble, between 1998 and 2002.   In this article, we focus solely on the analysis, which we have carried out anew.  We begin, in \secabbr \ref{sec:n_freq_fitting}, with a description of the determination of the neutron resonant frequency via the Ramsey method of separated oscillatory fields in conjunction with the cycle-by-cycle corrections from the mercury co-magnetometer.

A number of cuts were applied to the data.  These are discussed in \secabbr \ref{sec:cuts}.

\begin{figure} [ht]
\begin{center}
\resizebox{0.5\textwidth}{!}{
\includegraphics{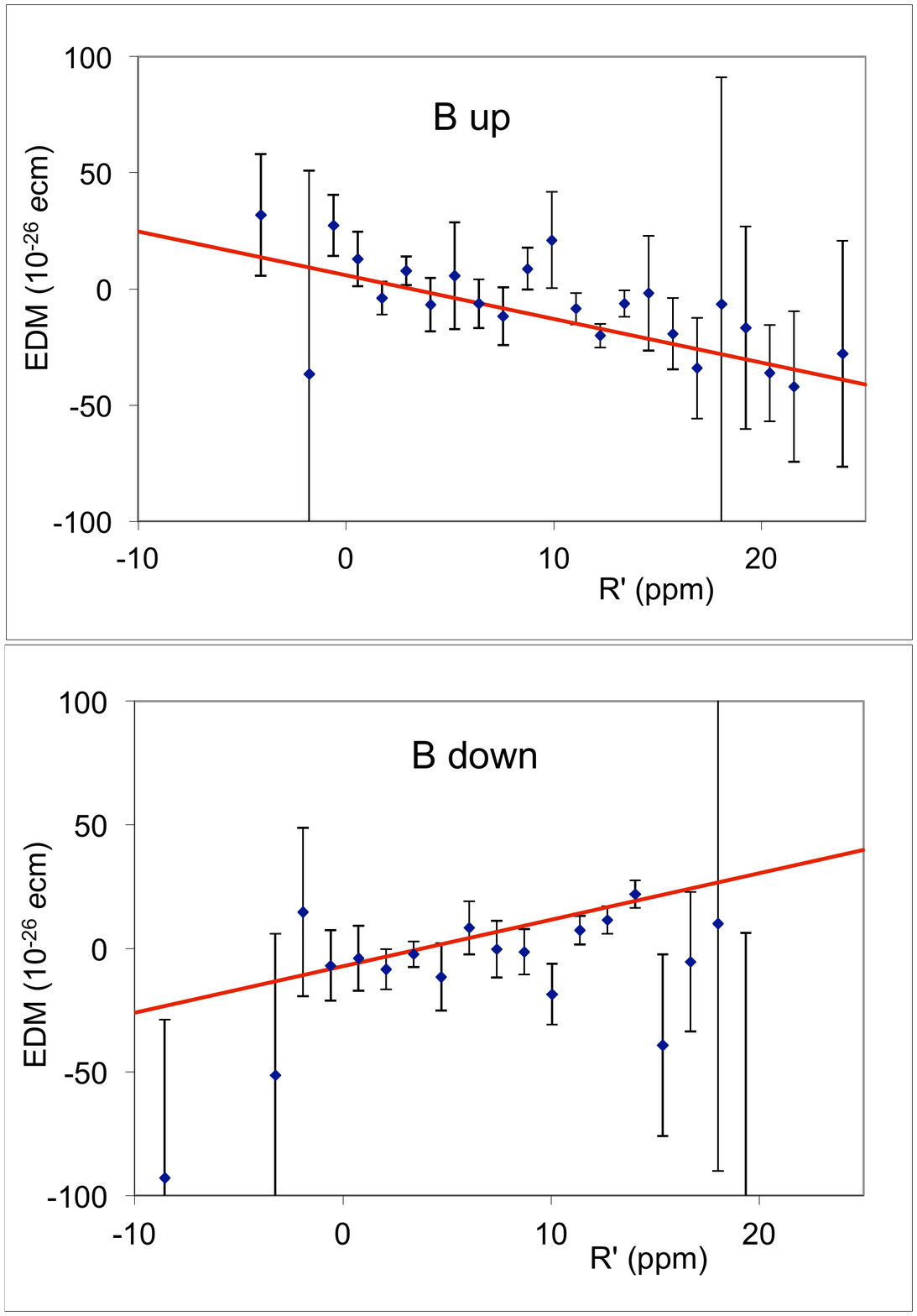}}
\end{center}
\caption{(Color online) From \cite{Baker_2006}.  Measured EDM (binned data) as a function of the relative frequency shift of neutrons and Hg.  The solid red line is a linear best fit.  
}
\label{fig:edm_vs_R}
\end{figure}

For each data-taking run, of typically 1-2 days' duration, a value of the EDM was determined from the slope of a linear fit of the (field-drift corrected) neutron frequency versus applied electric field $E$.  However, in the presence of a magnetic-field gradient, the mercury (and, to a far lesser extent, neutron) frequency acquires a component that is linear in the applied electric field \cite{Pendlebury_2004}, and thus mimics the signal for an EDM.  We have no direct measurement of the applied $B$-field gradient, but since the neutrons have a lower center of mass than the (thermal) mercury atoms, any change in the gradient results in a change in the neutron-to-mercury frequency ratio.  A linear dependence of the measured apparent EDM as a function of this ratio therefore emerges, as shown in \figabbr 2 of \cite{Baker_2006} and reproduced here in \figabbr \ref{fig:edm_vs_R}, with a positive (negative) slope for a downwards (upwards) direction of the applied magnetic field $\vec{B}$.   We discuss this further in \secabbr \ref{sec:false_edm} below.

The offsets and profiles of these lines can be affected by a number of factors, including nonuniform and/or horizontal magnetic-field gradients, the gravitationally enhanced spectrum-dependent depolarization of the ultracold neutrons (UCN), and even the rotation of the Earth \cite{Lamoreaux_2007}.  These are discussed in \secabbr \ref{sec:false_edm_mods}.

In order to measure and compensate for such effects, measurements were made in an auxiliary bottle that used a lid of adjustable height, so that the neutron and mercury frequencies could be determined as a function of height for a given $B$-field configuration.  These measurements are described in \secabbr \ref{sec:aux_meas}.

In \secabbr \ref{sec:globalfit}, we describe how all of these measurements are brought together in a global fit, in order to determine the necessary corrections to the initial EDM estimate obtained from the fit to the two crossing lines.  

In \secabbr \ref{sec:other_syst} we discuss further possible contributions to systematic errors.

We close, in \secabbr \ref{sec:final_result}, with a final summary and conclusion.

For convenience in what follows, and following the convention of \cite{Baker_2006}, we define
\begin{equation}
\label{eqn:Ra}
R_a = \left|\frac{\nu_\mathrm{n}}{\nu_{\mathrm{Hg}}} \frac{\gamma_{\mathrm{Hg}}}{\gamma_\mathrm{n}}\right|,
\end{equation}
where $\nu$, $\gamma$ are the respective frequencies and gyromagnetic ratios of the two species (neutrons and Hg), and we also introduce 
\begin{equation}
\label{Rprime}
R^\prime = R_a-1.
\end{equation}
In this context, $R^\prime$ is usually specified in parts per million (ppm). Throughout this analysis we have used values of $\gamma_\mathrm{Hg} = 7.590118(13)$\,Hz/$\mu$T and $\gamma_\mathrm{n} = -29.1646943(69)$\,Hz/$\mu$T, based on the measurements of Cagnac \cite{cagnac61} and of Greene et al.\ \cite{Greene_1979}.  Both of these measurements were made relative to the shielded proton gyromagnetic ratio in pure water, for which the most recently measured value is $\gamma_\mathrm{p}^\prime =  42.5763866(10)$\,Hz/$\mu$T \cite{Mohr_2012}.  From these results, we observe that  $R^\prime = 0$  when $\nu_\mathrm{n}/\nu_\mathrm{Hg}$ = 3.8424560(66), which represents an accuracy of 1.73\,ppm.  We incorporate the recent measurement of $\gamma_\mathrm{n}/\gamma_\mathrm{Hg} = -3.8424574(30)$ by Afach et al.\ \cite{Afach_2014a} into the later stages of our analysis.

\section{Neutron frequency determination}

\label{sec:n_freq_fitting}

The neutron frequency fitting procedure is
described in \cite{Green_1998} and \cite{Baker_2014}. For any measurement cycle (equivalently referred to as a ``batch'' in \cite{Baker_2014}), the first-order estimate $\nu_\mathrm{n,Hg}$ of the neutron resonant frequency is determined from the measured mercury precession frequency $\nu_\mathrm{Hg}$ by
\begin{equation}
\nu_\mathrm{n,Hg} = \left|\frac{\gamma_\mathrm{n}}{\gamma_\mathrm{Hg}}\right| \nu_\mathrm{Hg}.
\end{equation}
The Ramsey lineshape may then be written as 
\begin{equation}
N=\overline{N}\left[ 1\mp \alpha \cos \left( {\frac{\delta \nu }{\Delta \nu }%
}\pi -\phi \right) \right] , 
 \label{eqn: Ramsey with phase}
\end{equation}
where $\delta \nu =\nu_\mathrm{n,Hg}-\nu _1$ is the difference between the first-order resonant-frequency estimate $\nu_\mathrm{n,Hg}$ and the frequency $\nu_1$ of the applied r.f.\ pulses.  The linewidth $\Delta \nu$ is the width at half height of the central fringe, and is given by
\begin{equation}
\label{eqn:linewidth}
\Delta \nu = \frac{1}{2 T +8t/\pi} \approx \frac{1}{2T},
\end{equation}
where $T$ is the free precession time and $t$ is the duration of each of the $\pi/2$ spin-flip pulses.  The polarization $\alpha$ is related to the fringe maximum and minimum $N_\mathrm{max}, N_\mathrm{min}$ as 
\begin{equation}
\alpha = \frac{N_\mathrm{max}-N_\mathrm{min}}{N_\mathrm{max}+N_\mathrm{min}}.
\end{equation}
The phase 
\begin{equation}
\phi =\frac{\nu_\mathrm{n,Hg}-\nu _0}{\Delta \nu }\pi 
\end{equation}
incorporates the difference between the true resonant frequency $\nu_0$ and the first-order estimate $\nu_\mathrm{n,Hg}$ derived from the mercury. This difference arises for several reasons. First,
there is the inherent uncertainty in the ratio $\gamma _\mathrm{n}/\gamma _{\mathrm{Hg}}$ as discussed above.  Second, and as also discussed above, $\phi$ alters because the neutrons were of such low energy that they
 populated preferentially the lower portion of the storage
volume, whereas the mercury sampled the entire volume uniformly; there was a
difference in height of the centers of gravity of the two systems of a few mm. In the presence of a vertical gradient $\dBzdz$ in the magnetic field, the average
field sampled by the mercury would have been slightly different to that sampled by
the neutrons.  The normalized frequency ratio is therefore related to this height difference $\Delta h$ by
\begin{equation}
\label{eqn:R_delta_h}
R^\prime = \pm \Delta h \frac{\dBzdz}{B_0},
 \end{equation}
where the + sign corresponds to $\vecB$\ downwards.

Third, as will be discussed below, gradients in horizontal components of the magnetic field can increase the neutron
precession frequency while leaving the mercury unaffected. The rotation of the Earth also has a part to play, as do minor effects such as light-induced frequency shifts in the mercury.

These factors, which remained essentially constant throughout each run,  caused the departure of $R_a$ (\eqnabbr \ref{eqn:Ra}) from unity. There are, in
addition, two factors that could provide extremely small cycle-to-cycle
variations in $\phi $. These are the small mismatch in temporal overlap of
the mercury and neutron frequency measurements, as discussed in \secabbr \ref
{sec:fieldjumps}, and a genuine neutron EDM or some effect that mimics it.

The HV was applied in a simple pattern of alternating sign, with an additional four cycles at $E=0$ between each 16 cycles of HV at a given polarity.  The inital polarity for each run was chosen randomly. In principle, one could treat each $E$ value within a run as a distinct data set, with its own independent Ramsey-curve fit.  Our approach instead has been to fit a single Ramsey curve to each entire run, with the frequency shifts (which are very small perturbations, well within the noise) then being analyzed on a cycle-by-cycle basis. This has the benefit of allowing analysis of runs containing insufficient data within each $E$ group for a reliable Ramsey fit.  

While the polarity of the $E$ field could be switched automatically under computer control, changing the direction of the magnetic field was a much more onerous procedure.  This was normally undertaken every few weeks.

\begin{figure} [ht]
\begin{center}
\resizebox{0.5\textwidth}{!}{
\includegraphics{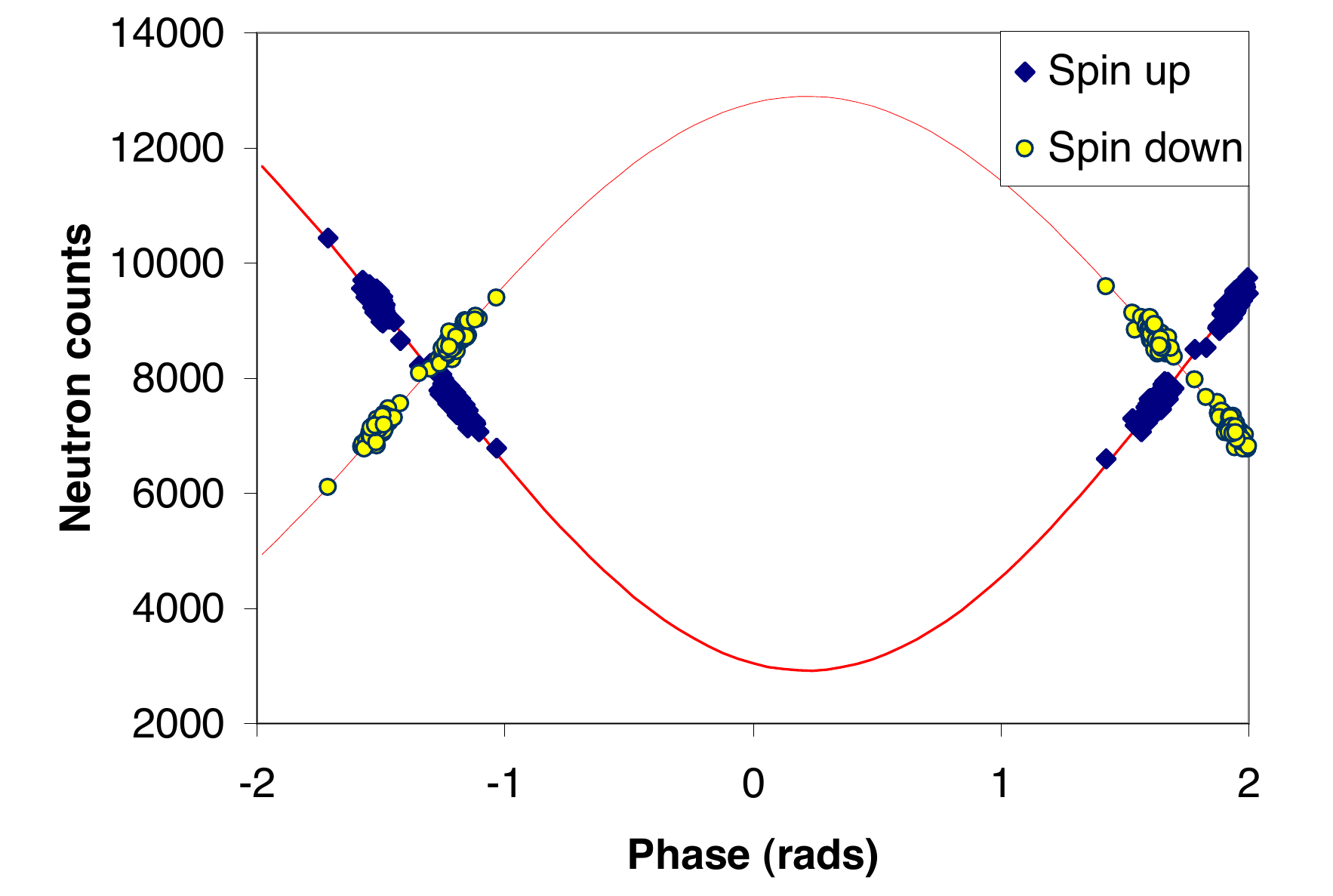}}
\end{center}
\caption{(Color online) From \cite{Baker_2014}.  Spin-up and spin-down neutron counts for a single run fitted to the Ramsey curve (\eqnabbr \ref{eqn: Ramsey with phase}).}
\label{fig:ramsey_fit}
\end{figure}

For each run, typically consisting of several hundred measurement cycles,  
the neutron counts were fitted to \eqnabbr (\ref{eqn: Ramsey with phase}), as shown in \figabbr \ref{fig:ramsey_fit}, to
provide values of $\overline{N_{s}}$, $\alpha _{s}$ and an average value of
the phase $\phi _{s}$ for each of the two spin states $s$.  (For maximum sensitivity, measurements were taken repeatedly at four working points, close to the half-height of the central Ramsey valley.) For the first
iteration of each of these two fits, the uncertainty $\sigma _{y}$ allocated
to each data point was simply the square root of the number of neutrons
counted in that particular spin state. The uncertainty in the fitted mercury frequency, which resulted in an
uncertainty $\sigma _{x}$ in the $x$ coordinate of each data point, was
incorporated by calculating an ``indirect'' error $\sigma _{I}$ from the
slope of each first-iteration fitted curve, 
\begin{equation}
\sigma _{I}=\frac{dy}{dx}\sigma _{x},
\end{equation}
and adding it in quadrature to the original error bar to obtain a new
uncertainty 
\begin{equation}
\sigma _{\mathrm{tot}}=\left( \sigma _{y}^{2}+\sigma _{I}^{2}\right) ^{1/2}
\end{equation}
for use in a second iteration of the fit. This correction was, however,
generally negligible, as the mercury usually measured the magnetic field with much
higher precision than did the neutrons.  

An overall average (weighted as the fit uncertainties) $\bar{\phi}$ of the two phases $\phi _{s}$ was
calculated, giving a common phase offset for the entire run and for
both spin states. \eqnabbr (\ref{eqn: Ramsey with phase}) was then inverted
to yield an individual phase shift $\delta \phi _{s,i}$ for each data point $%
i$ and for each spin state: 
\begin{equation}
\delta \phi _{s,i}={\frac{\delta \nu }{\Delta \nu }}\pi -\bar{\phi}-\arccos
\left[ \frac{N_{s,i}-\bar{N}_{s}}{\bar{N}_{s}\alpha _{s}}\right].
\label{eqn: individual phase shifts}
\end{equation}
A phase shift $\delta \phi _{i}$ was calculated for each data point by
averaging the $\delta \phi _{s,i}$ over the two spin states. This averaging
helped to remove non-statistical fluctuations in the total neutron flux,
which could, exceptionally, vary by a percent or two during a run.

Any shift $\delta \nu _{0i}$ in the neutron resonant frequency from the
value $\nu _0$ predicted by the fitted Ramsey curve should be proportional
to this phase shift $\delta \phi _i$: 
\begin{equation}
\delta \nu _{0i}=\delta \phi _i\cdot \frac{\Delta \nu }\pi .
\end{equation}
In the presence of an EDM, this frequency shift should be directly
proportional to the strength of the electric field. A straightforward linear least-squares
 fit of the frequency shifts $\delta \nu _{0i}$ as a function of the
applied electric field $\vec{E}$ yielded a value for the measured EDM $d_\mathrm{meas}$,
with its associated uncertainty, for each run.  

We have recently carried out detailed modeling of our storage-chamber geometry using the Opera \cite{opera} finite-element analysis package, and we find that the average electric field within the volume is 1.1\% lower than the nominal value determined from the applied voltage divided by the separation between the electrodes. .

For completeness, we note that the statistical uncertainty to be expected from an EDM measurement based upon the Ramsey technique is \cite{Baker_2014} 
\begin{equation}
\sigma_d\approx \frac{\hbar}{2\alpha E T \sqrt{N}}.
\label{eqn:edm_uncert}
\end{equation}

Care was taken to verify that the analysis delivered the correct sign of EDM: in particular, with $\vec{E}$ and $\vec{B}$ fields parallel, an increase in neutron precession frequency characterizes a negative EDM.


\section{Applied cuts}
\label{sec:cuts}

In this \secabbr we list all cuts applied to the data.  No systematic dependence of the EDM signal upon any of these cuts was observed.  After all of these following cuts were applied, 545 runs containing 175,217 measurement cycles and $2.5\times10^9$ neutrons remained.

\subsection{Manual cuts}

After inspection, 10 runs were rejected for various reasons relating to the failure of hardware components such as amplifiers, valves, the HV supply and so on.    Some individual cycles within each run were also cut for similar reasons: the majority of these were due to problems with the delivery of neutrons.    A further 26 runs were rejected because they had 19 or fewer measurement cycles, which is not sufficient for the HV polarity reversal required for an EDM measurement. 

These manual cuts removed about 4\% of the available data, after which the above-mentioned 545 runs remained. 

\subsection{Sub-runs}

During part of the data taking, an adjacent experiment was using a superconducting magnet.  When turned on or off, $R_a$ changed by typically 2-3\,ppm.  The eleven measurement runs during which this occurred were therefore split into sub-runs for separate analysis.

\subsection{Mercury $\chi^2/\nu$ cut}

\begin{figure} [ht]
\begin{center}
\resizebox{0.5\textwidth}{!}{\includegraphics*{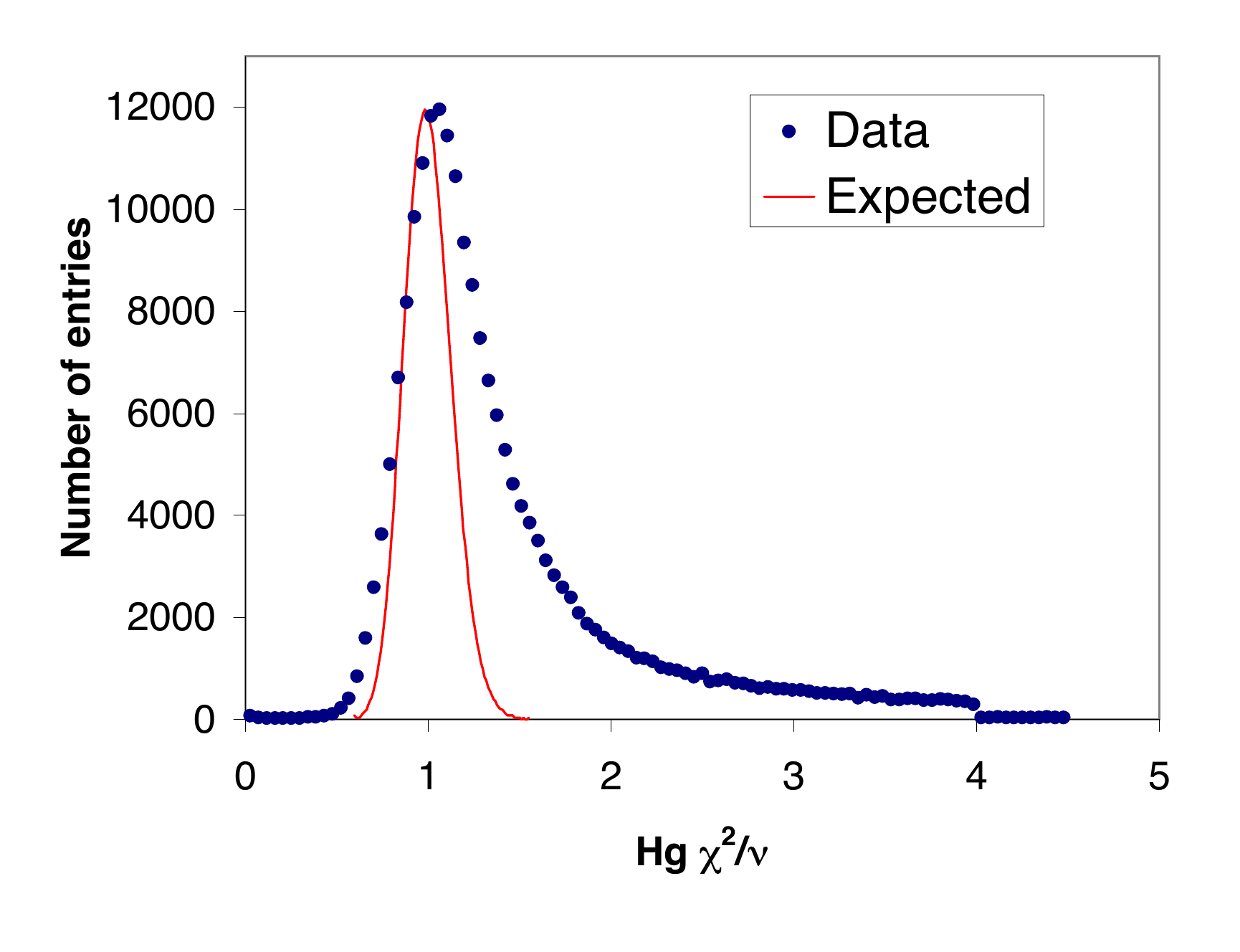}}
\end{center}
\caption{(Color online) From \cite{Baker_2014}. The distribution of $\chi ^2/\nu $
from mercury frequency fits.  See text for details.}
\label{fig:Hg_fit_chisq}
\end{figure}

The $\chinu$ distribution from the cycle-by-cycle online Hg frequency fitting is shown in \figabbr \ref{fig:Hg_fit_chisq}, which is reproduced from \figabbr 10 of \cite{Baker_2014}.  The red ``expected'' distribution curve is based upon the model of a perfectly constant frequency, whereas in fact we do expect some slow variation over time within a cycle.  That being the case, a large $\chinu$ is not necessarily indicative of a problem with the frequency measurement, and we therefore retain a generous portion of this distribution beyond the vicinity of the peak.  A cut was made at $\chinu$ = 3, by which time the tail is fairly flat.  

This cut removed nearly 7\% of the remaining data.

We note in passing the discontinuity at $\chinu$ = 4.  This arises because, beyond this point, the online fitting procedure attempted to correct for potential hardware errors (e.g.\ gain saturation or a missed ADC reading), as discussed in \cite{Baker_2014}.   

\subsection{Mercury frequency uncertainty}

\begin{figure} [ht]
\begin{center}
\resizebox{0.5\textwidth}{!}{\includegraphics*{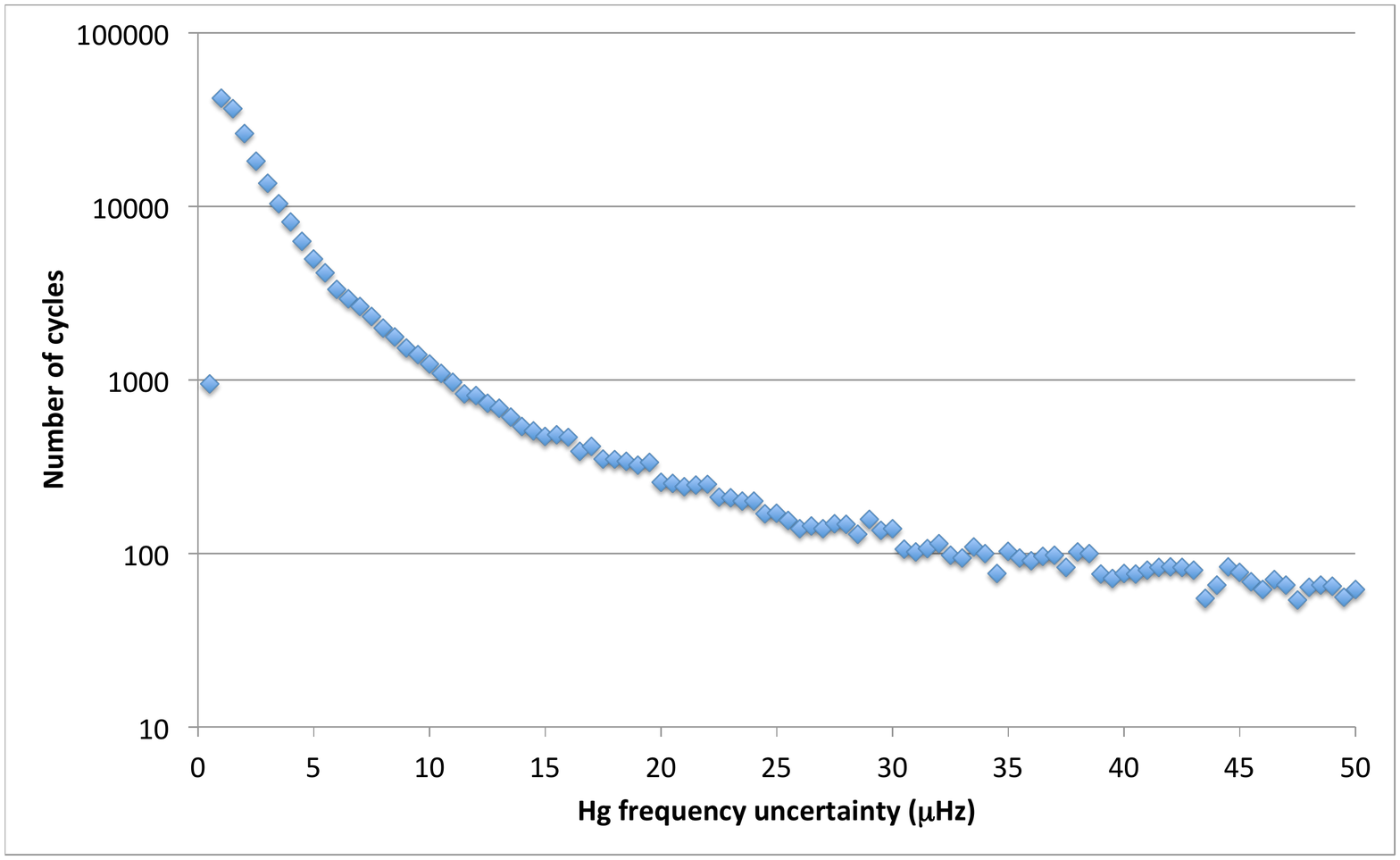}}
\end{center}
\caption{(Color online) The distribution of uncertainties from mercury frequency fits.  See text for details.}
\label{fig:Hg_errors}
\end{figure}

The fitted Hg frequency sometimes had a large uncertainty, particularly if the depolarization time was short.  The distribution of these uncertainties is shown in \figabbr 13 of \cite{Baker_2014}, and again here on a semi-log plot in \figabbr \ref{fig:Hg_errors}.  A typical value is 1-2\,$\mu$Hz, which (when scaled with $\gamma_n/\gamma_\mathrm{Hg}$) corresponds to approximately a factor of 5 better than the  typical inherent neutron frequency uncertainty from counting statistics.   A cut was made at 25\,$\mu$Hz, at which point the relationship is approximately inverted, with the mercury uncertainty entirely dominating the frequency-ratio measurement.  Beyond this, the data are so imprecise that they make essentially no contribution at all.

This cut removed 4\% of the remaining data.

\subsection{Magnetic-field jumps}
\label{sec:fieldjumps}

The distribution of Hg frequency jumps, i.e.\ the difference in Hg frequency between a given cycle and the previous cycle, is shown 
in \figabbr 14 of \cite{Baker_2014}.  There are broad tails due to occasional sudden changes in field, for example due to the movement of an overhead crane or to a mechanical disturbance to the $\mu$-metal shields.

The mercury and the neutron frequency measurements do not have perfect temporal overlap.  As discussed in \cite{Baker_2014}, the Hg frequency was determined by fitting a 15-second averaging period at either end of the cycle in order to determine the phase, and hence the integrated phase difference accumulated over the time between them.  The corresponding phases for the neutrons, on the other hand, are determined by the 2-second spin-flip pulses at either end.  
The field jumps can occur at any point during the 220\,s measurement cycle.  If they occur outside the Ramsey sequence, they are of little concern.  If they occur between the 15\,s windows, they can be regarded as appropriately compensated.  That leaves a potential 30\,s period during which there is a risk of incomplete compensation, i.e.\ 1/7 of such jumps can be expected to affect the measurement to a greater or lesser extent.  On average, these potentially risky field jumps would occur halfway through the 15\,s window, i.e.\ about 1/20 of the way through the Ramsey measurement.  The mercury  frequency-jump distribution was truncated at $\pm 60$\,$\mu$Hz, corresponding to a change in neutron frequency of 230\,$\mu$Hz over this 1/20 of the Ramsey period.  When averaged over the entire Ramsey period, the neutrons would therefore see a frequency shift of up to  11\,$\mu$Hz, or 0.4\,ppm, that would not be compensated properly by the mercury.  Including the aforementioned 1/7 probability, this corresponds (even at the extremes of the frequency-jump distribution) to a potential error in $R^\prime$ of 0.06\,ppm on the rare cycles within which such jumps occur, to be compared with a typical statistical uncertainty on the neutron frequency of about 0.7\,ppm.

The frequency-jump cut removed 3\% of the remaining data.  

\subsection{First-cycle cut}

The first cycle of any run is different from any of the others, as the neutron trap and guides are initially empty; for other cycles there is likely to be some remnant population from the previous cycle.  In consequence the first cycle often has an anomalously low total neutron count.  

This cut removed 0.2\% of the remaining data.

\subsection{Ramsey residuals outlier cut}

\begin{figure} [ht]
\begin{center}
\resizebox{0.5\textwidth}{!}{
\includegraphics{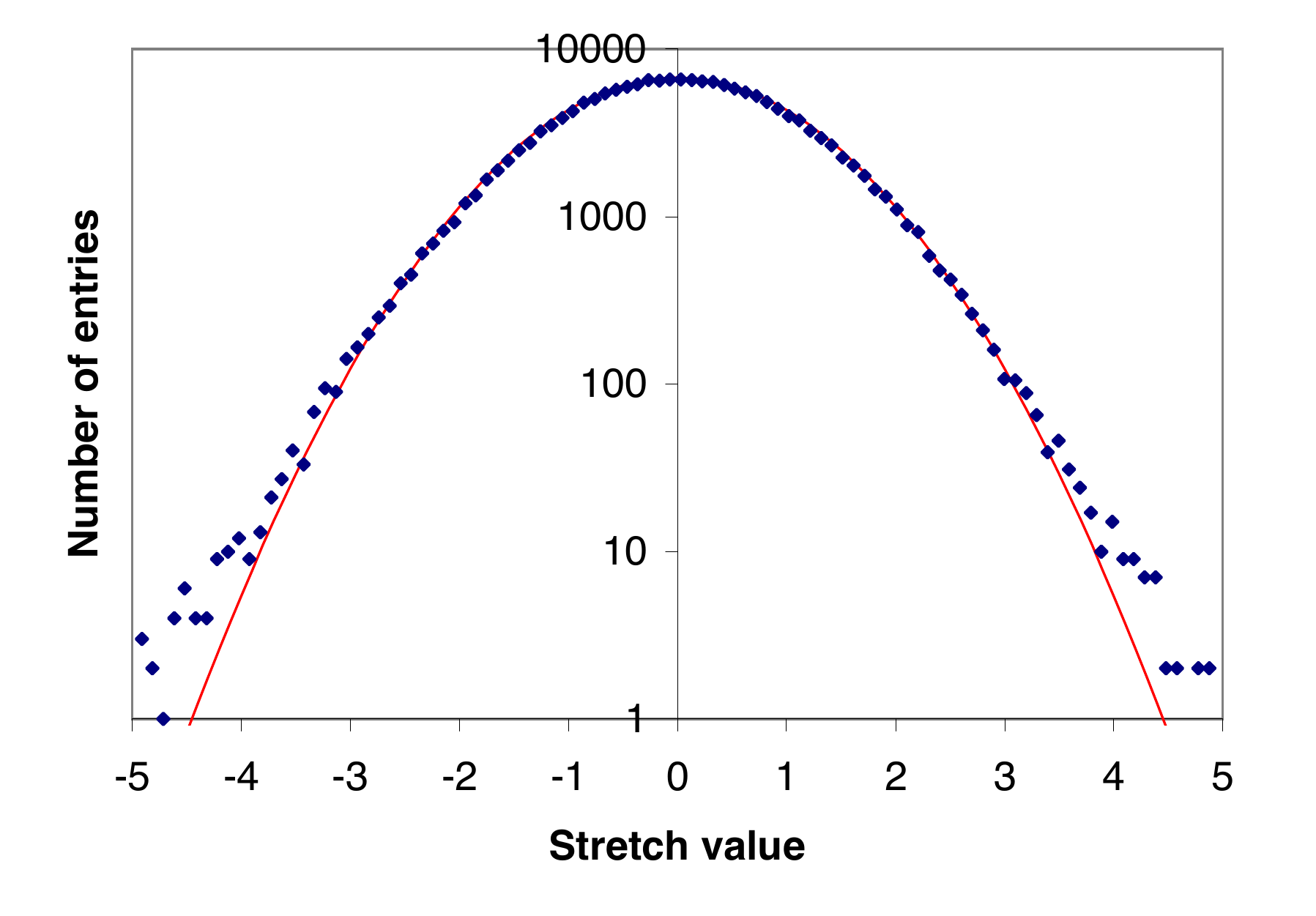}}
\end{center}
\caption{(Color online)  From \cite{Baker_2014}.  Distribution of stretch values from the fits to the Ramsey curve }
\label{fig:ramsey_residuals}
\end{figure}

The data points of \figabbr \ref{fig:ramsey_residuals}, reproduced from \figabbr 5 of \cite{Baker_2014}, show the distribution of stretch values $r_i$ of the fits to the Ramsey curve: 
\begin{equation}
\label{eqn:ramseystretch}
r_i = \frac{\left(\nu_i -\nu_{R_i}\right)}{\sigma_i},
\end{equation} 
where $\nu_i$ is the calculated frequency of the $i$th cycle, $\sigma_i$ is its uncertainty and $\nu_{R_i}$ is the expected frequency for that cycle as defined by the Hg frequency, the applied r.f.\ and the Ramsey curve function.  Ideally, and in the absence of any EDM-like signals, this distribution would be expected to be a Gaussian of unit width.  The continuous line is a Gaussian of width 1.06.   The true distribution departs from this Gaussian at about 4$\sigma$.  The few points lying outside this range tend to be associated with runs that have other known problems, for example with intermittent failure of the neutron delivery system.  A cut was therefore made at $\pm 4\sigma$.  Because of the symmetric way in which the data are taken, cutting the tails from this distribution cannot of itself induce a false EDM signal.

This cut removed a further 0.05\% of the remaining data. 

\subsection{High voltage}

\begin{figure} [ht]
\begin{center}
\resizebox{0.5\textwidth}{!}{
\includegraphics{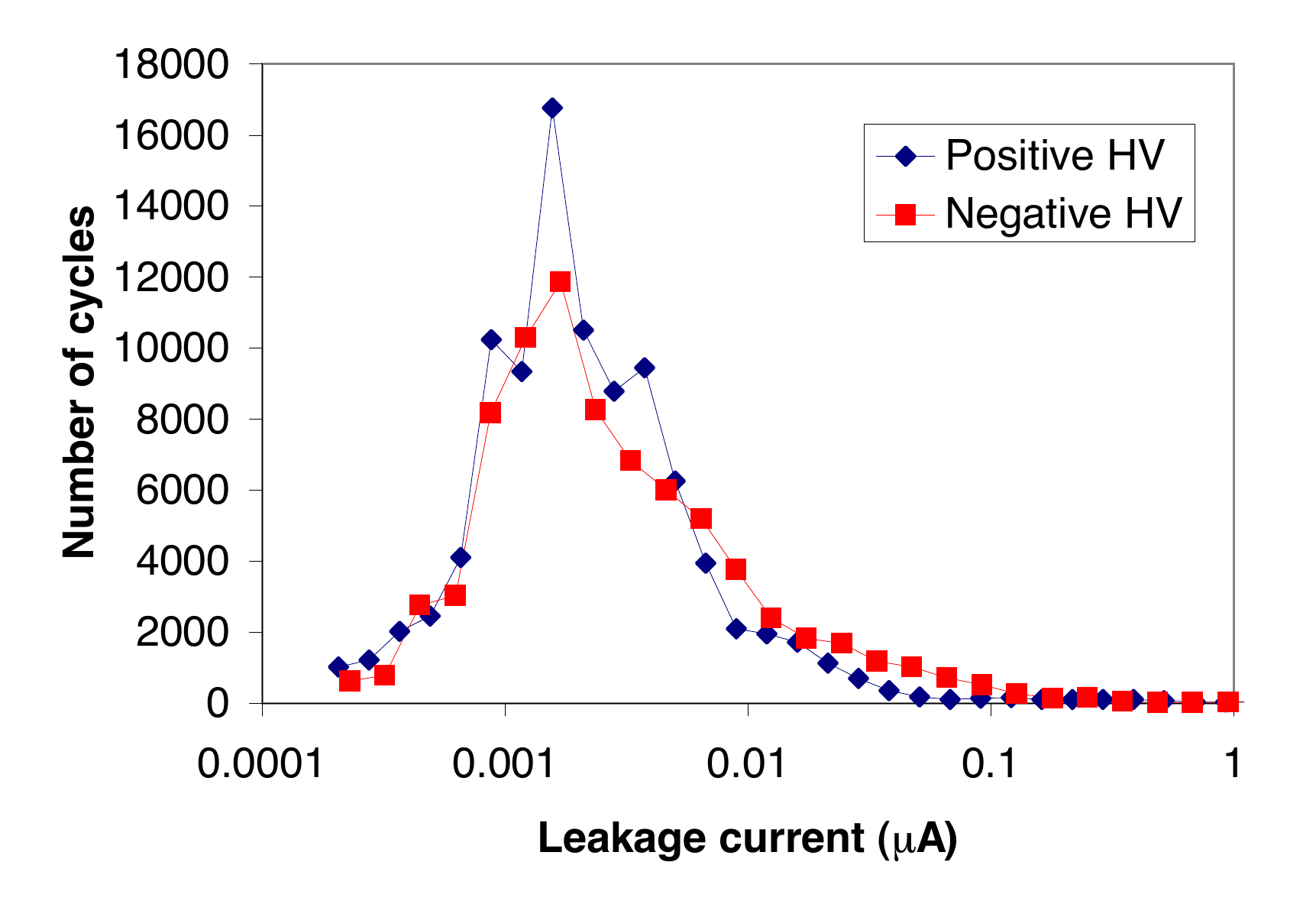}}
\end{center}
\caption{(Color online) From \cite{Baker_2014}. Distribution of the cycle-by-cycle average leakage currents $I$}
\label{fig:leakage_current_distrib}
\end{figure}

\begin{figure} [ht]
\begin{center}
\resizebox{0.5\textwidth}{!}{
\includegraphics{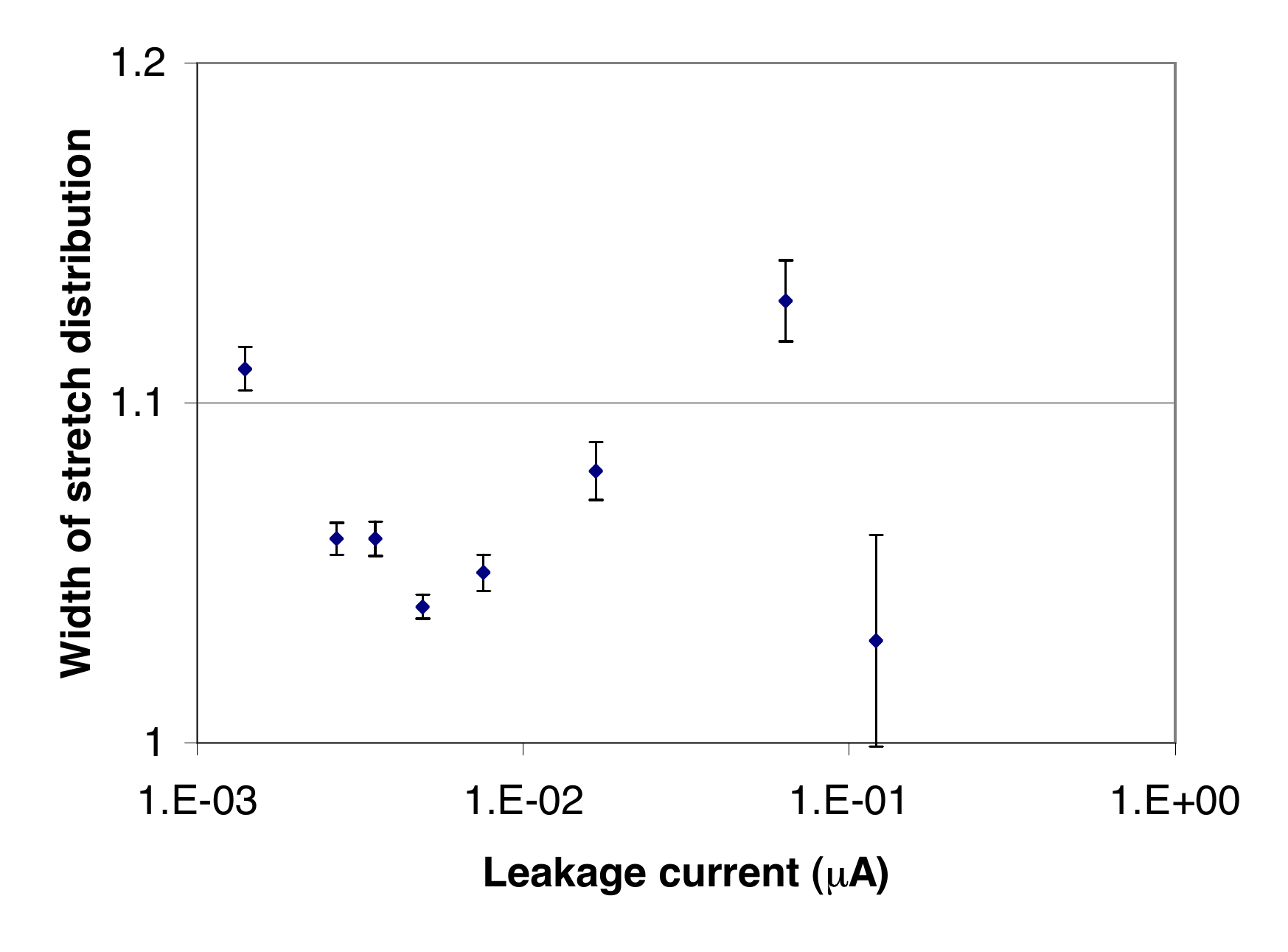}}
\end{center}
\caption{(Color online) Departure from the Ramsey curve, as measured by the width of the stretch value distribution (as in \figabbr \ref{fig:ramsey_residuals}), as a function of the leakage current $I$.  The widths are constant to within a few percent, and there is no overall trend.}
\label{Fig:stretch_vs_I}
\end{figure}

A single measurement cycle with a leakage current in excess of 60\,$\mu$A was removed from the data set.  All remaining measurement cycles had leakage currents below 10\,$\mu$A, with the great majority being of the order of a few nA; the distribution for both polarities is shown in \figabbr \ref{fig:leakage_current_distrib}, which reproduces \figabbr 20 of \cite{Baker_2014}.   No further cuts relating to the HV were made: the mercury magnetometer is relied upon to compensate for any residual effects
from this source. Under the assumption that leakage currents would run along preferred established paths -- rather than averaging out over different paths -- then if such leakage currents were to generate EDM signals, measurement cycles with high leakage currents would be expected to show a greater-than-average departure from the Ramsey curve, and thus to have high Ramsey-residual stretch values (\eqnabbr \ref{eqn:ramseystretch}).  In order to quantify this, the data were binned by leakage current, and the distributions of stretch values within each bin were fitted to Gaussians.  The widths of these Gaussians are plotted against leakage current in \figabbr \ref{Fig:stretch_vs_I}.  No consistent trend is visible.  Furthermore, after processing, no dependence of the measured EDM on leakage current was observed, as discussed in \secabbr \ref{sec:syst_leakage}.

\subsection{Frequency ratio}

\begin{figure}[ht]
  \begin{center}
    \resizebox{0.5\textwidth}{!}{\includegraphics
    [clip=true, viewport = 50 430 550 760]{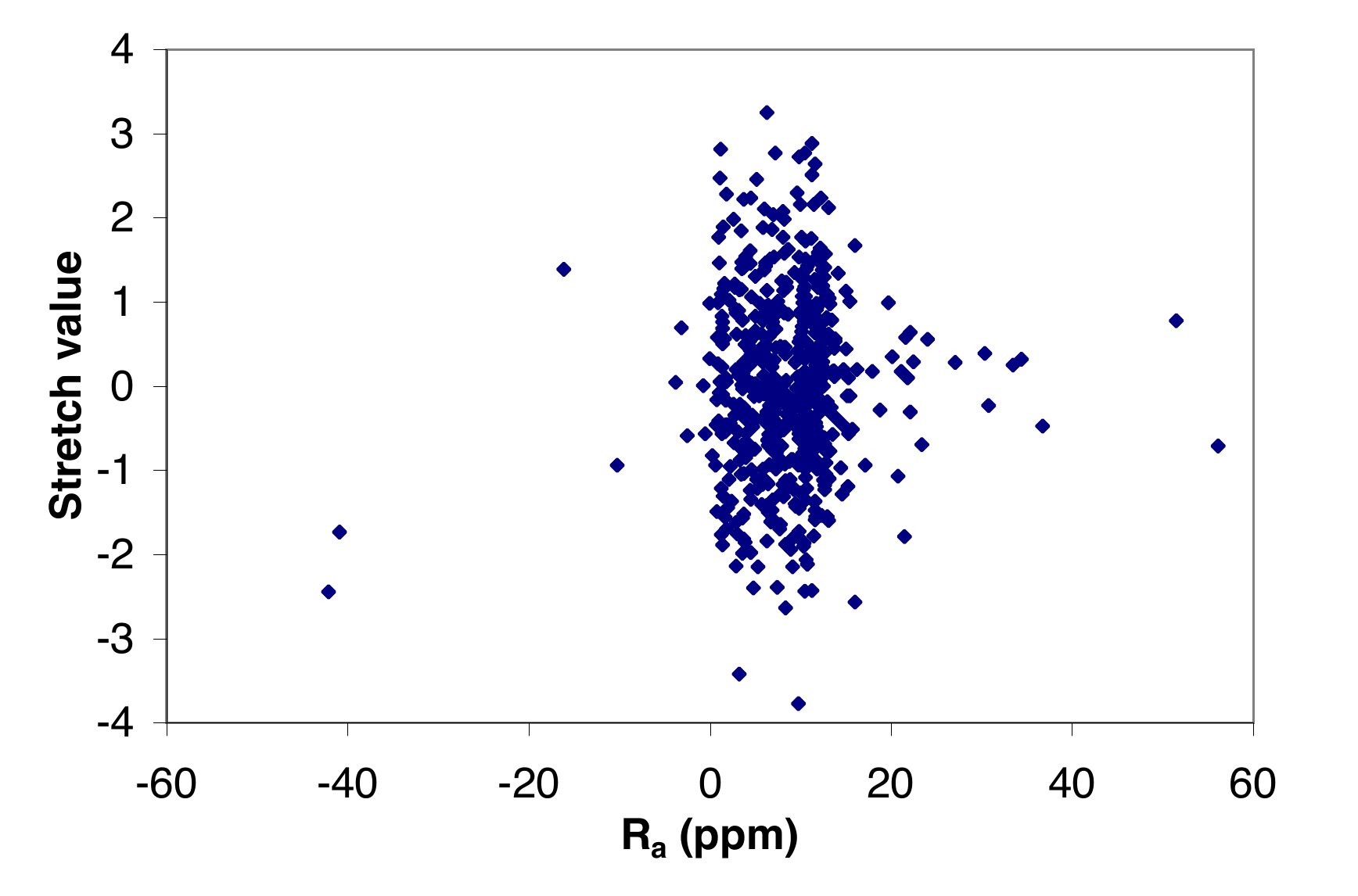}}
    \caption{(Color online)  Measured EDM stretch value as a function of the frequency ratio $R_a$}
    \label{fig:dstretch_vs_R}
  \end{center}
\end{figure}

Measurements were undertaken with a range of different applied $B$-field gradients $\dBzdz$ by pre-adjusting currents in field-trimming coils.  (Initially, this arose through trial-and-error as the system was optimized; we then settled upon a more consistent configuration for each $B_0$ direction, with a few runs later on used to explore the effects of deliberately large gradients.)    As discussed in \secabbr \ref{sec:false_edm}, these field gradients induced false EDM signals.  Any nonlinearity  in this effect would appear as a systematic departure of the points in \figabbr \ref{fig:edm_vs_R} from their fitted lines. We define the  stretch values
\begin{equation}
\label{eqn:gp_stretch}
z_i = \frac{d_{\mathrm{meas}_i}-d_{f_i}}{\sigma_i},
\end{equation}
where $d_{\mathrm{meas}_i}\pm \sigma_i$ is the measured EDM for run $i$, and $d_{f_i}$ is the corresponding value from the fitted line.  The distribution of these stretch values is shown in 
\figabbr \ref{fig:dstretch_vs_R}.  There being no evidence of any correlation, and none expected, it was decided not to impose any restriction upon the range of $R_a$.

\section{False EDM signals}
\label{sec:false_edm}

\subsection{Introduction}

\begin{figure}[ht]
  \begin{center}
    \resizebox{0.5\textwidth}{!}{\includegraphics   
 {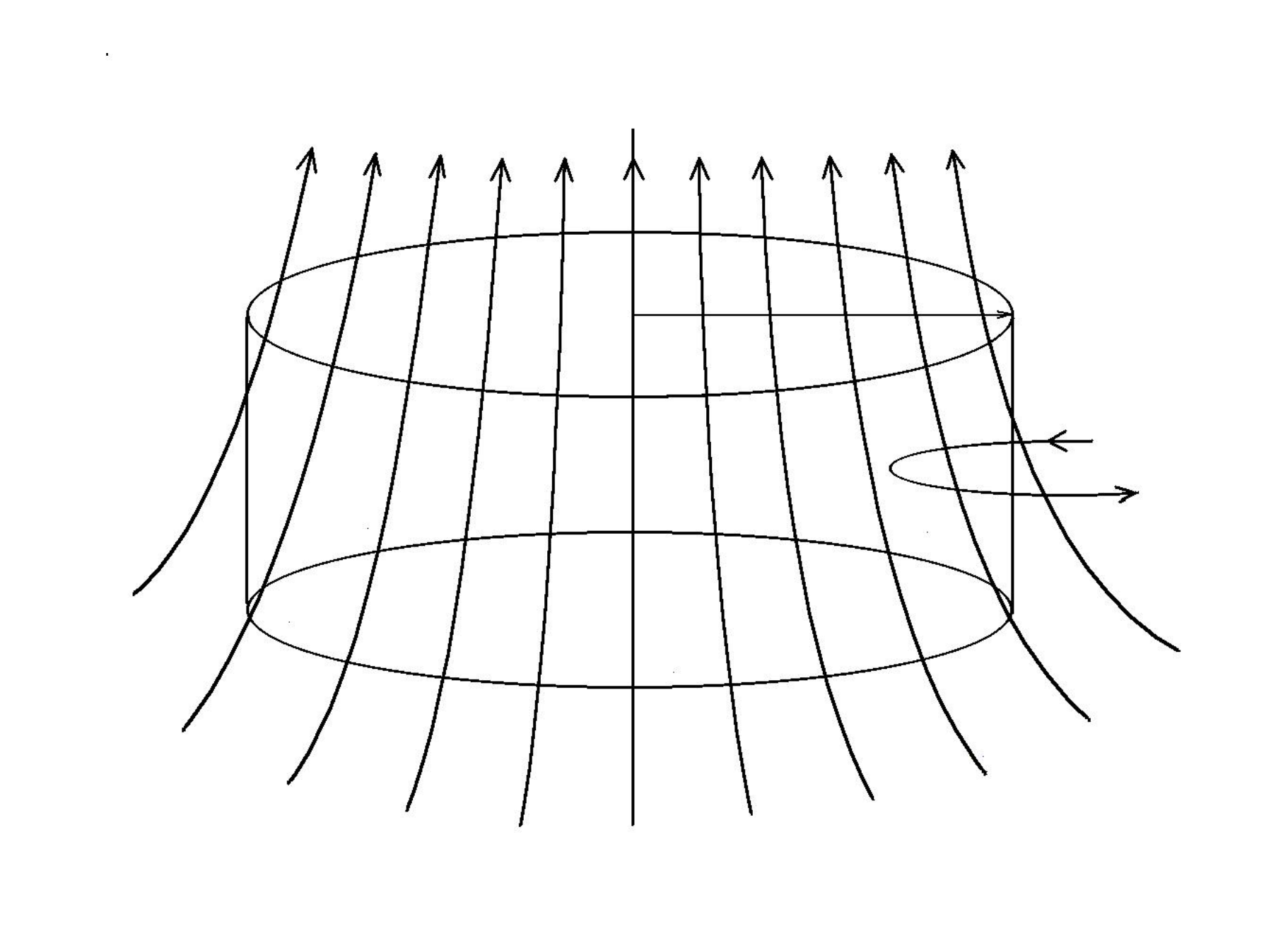}}
    \caption{From \cite{Pendlebury_2004}.  Showing the shape of the $\vecB$ field lines, when there is a positive gradient  $\dBzdz$, shown in relation to an outline of the trap used to store $^{199}$Hg atoms and ultracold neutrons (UCN). Additional fields having lines that both enter and leave through the sidewalls, like the one on the right-hand side, do not affect the false EDM signals that are generated.
}
    \label{fig:geophase_Bfield_trap}
  \end{center}
\end{figure}

A false EDM signal $\dnf$ arises when the trapped particles experience a gradient $\dBzdz$ in the presence of the electric field $E$ \cite{Pendlebury_2004}.  The effect is easiest to understand in the case of an azimuthally-symmetric field with a vertical gradient  $\dBzdz$, i.e., a slightly trumpet-shaped field, as shown in \figabbr \ref{fig:geophase_Bfield_trap}.  

\begin{figure}[ht]
  \begin{center}
\resizebox*{0.5\textwidth}{!}{\includegraphics 
 {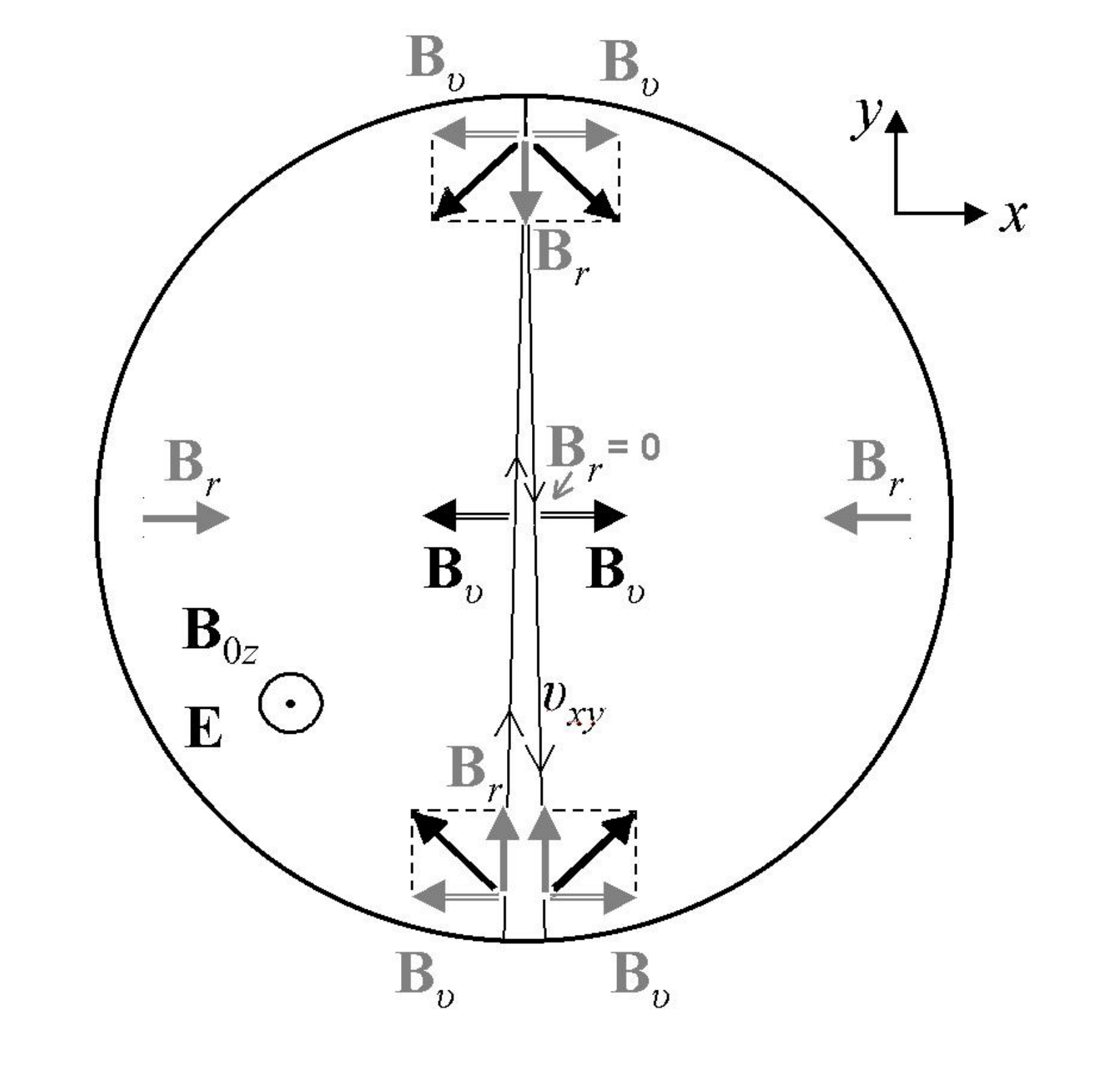}}
    \caption{From \cite{Pendlebury_2004}.  Showing the $B_{xy}$ fields (in black) seen by a particle going back and forth close to the $y$-axis. Going towards positive $y$, the $B_{xy}$ field rotates steadily anticlockwise by about 70$^\circ$ as drawn. The first reflection of the particle towards negative $y$ causes an instantaneous anticlockwise rotation by about 110$^\circ$ as drawn. The same two rotations occur on the path to, and at, the second reflection. The size of the rotations depends on the size of $B_{0r}/B_v$.
}
    \label{fig:geophase_effect}
  \end{center}
\end{figure}

Because $\vec{\nabla}\cdot\vec{B} = 0$, field lines either enter or emerge from the side walls of the cell, giving a radial field of strength that is proportional to the radius $r$:
\begin{equation}
B_r = -\frac{r}{2}\cdot\frac{\partial B_z}{\partial z}.
\end{equation}
Consider now a particle moving at speed $v$ that crosses the storage cell close to its diameter, as shown in \figabbr \ref{fig:geophase_effect}.  
As it travels through the electric field, the particle experiences in its own rest frame an
additional magnetic field  \cite{jackson_electrodynamics} 
\begin{equation}
\vec{B_v}\approx {-}\frac{\vec{v}\times \vec{E}}{c^{2}} 
\label{eqn:vxE}
\end{equation}
above and beyond the laboratory magnetic field $\vec{B}_{0}.$
At the start of its trajectory, just left of centre at the bottom, it is subject to the radial field $B_r$ as well as to the sideways $\vec{B}_v$ component, yielding a diagonal resultant.  As it traverses the trap, the $B_r$ component shrinks and then reverses direction, causing a smooth rotation of the net additional effective $B$ field.  Eventually, the particle reaches the far end of the trap, is reflected from the wall and begins its trajectory back.  However, the $\vec{B}_v$ component then faces in the opposite direction: therefore, after a discrete jump at the point of reflection, the additional net effective field component continues to rotate in the same direction.  The particle thus sees a rotating field in the $x$-$y$ plane, which, through the Ramsey-Bloch-Siegert \cite{Ramsey_1955,Bloch_1940} mechanism, `pulls' its resonant frequency away from the central value.  When averaged over both directions of a given trajectory, the net frequency shift is proportional to $E$, and it therefore mimics an EDM signal.

The size of this false-EDM effect depends upon the relative magnitudes of the orbital-trajectory frequency of the particle and its Larmor precession frequency.  The neutrons are in the adiabatic limit, where their rapidly precessing spins can follow the variations in field.  In that case, the false EDM is given by \cite{Pendlebury_2004}
\begin{equation}
\label{eqn:adiabatic_false_edm}
\delta \nu = \frac{v_{xy}^2}{8\pi B_0^2c^2}\frac{\partial B_z}{\partial z}E,
\end{equation}
where $v_{xy}$ is the average transverse particle velocity.  The origin of this shift is a geometric (or Berry's) phase \cite{Berry_1984}, and therefore -- as pointed out explicitly in \cite{Rebreyend_2015} -- is independent of the coupling to the magnetic field; the gyromagnetic ratio is absent from this equation.

The mercury lies in the non-adiabatic regime, and its frequency shift in such a field gradient is \cite{Pendlebury_2004}
\begin{equation}
\label{eqn:nonadiabatic_false_edm}
\delta\nu = \frac{\gamma^2r_B^2}{16\pi c^2}\frac{\partial B_z}{\partial z} E,
\end{equation}
where $r_B$ is the radius of the storage bottle. This has been independently measured \cite{Rebreyend_2015}.  A more general expression, valid for arbitrary magnetic fields in the non-adiabatic limit, was derived more recently \cite{Pignol_2012}:
\begin{equation}
\label{eqn:nonadiabatic_arb_field_false_edm}
\delta\nu = \frac{\gamma^2}{2\pi c^2}\left< xB_x + yB_y \right> E,
\end{equation}
where the brackets refer to the volume average over the (arbitrarily shaped) trap.

Since the mercury is used to compensate for shifts in the magnetic field, any EDM-like component contributing to the mercury frequency will affect the measurement of the neutron EDM.   We note that  \cite{Pignol_2012} contains a sign error in its \eqnabbr 20 and \figabbr 2, arising from this transferral of the mercury false EDM to the neutron signal.  

In this experiment, the contribution to $d_\mathrm{meas}$ of the false-EDM effect in the mercury is about 50 times larger than the geometric-phase induced false-EDM effect of the UCNs.

\subsection{False-EDM analysis: First iteration}

Combining \eqnabbr \ref{eqn:R_delta_h} with \eqnabbr \ref{eqn:nonadiabatic_false_edm}, the mercury's false-EDM contribution to $d_\mathrm{meas}$ is seen to be
\begin{equation}
\dnf = \pm\frac{\hbar}{8}|\gamma_\mathrm{n}\gamma_{\textit{Hg}}|\frac{r_B^2B_{0_z}}{\Delta h\,c^2}\cdot R^\prime = \pm k\cdot R^\prime,
\label{eqn:daf_Hgn}
\end{equation}
where $r_B$ is the trap radius and the + sign again corresponds to $\vecB$\ downwards.  The notation $d_\mathrm{n,Hg,f}$  is drawn from \eqnabbr 87 of \cite{Pendlebury_2004}.  It follows that we can write 
\begin{equation}
\label{eqn:d_meas}
d_\mathrm{meas} = d^\prime_\mathrm{n} + d_\mathrm{n,Hg,f} = d^\prime_\mathrm{n} \pm k\cdot(R^\prime-R^\prime_{0}),
\end{equation}
where $d^\prime_\mathrm{n}$ is the true $d_\mathrm{n}$ plus all other systematic effects discussed below, and $R^\prime_{0}$ is the value of $R^\prime$ where $\dBzdz = 0$.  \eqnabbr \ref{eqn:d_meas} defines two straight lines, one with a positive slope for $\vecB$\ down and one with a negative slope for $\vecB$\ up.  Na\"{i}vely, one would expect that the crossing point $(\Rx, d_\times)$ would occur at $\dBzdz = 0$, and would therefore provide an estimator of $d^\prime_\mathrm{n}$ free of $\dnf$.  In reality, as we discuss below, various effects can induce shifts in this crossing point, and we therefore must apply appropriate corrections.

 As discussed above,  \figabbr \ref{fig:edm_vs_R} shows the data (binned for clarity) for $d_\mathrm{meas}$ as a function of $R^\prime$ for each direction of $\vecB$.  The solid red lines represent a least-squares fit to all 545 of the (unbinned) run results, using as free parameters the two intercepts $c_1, c_2$ and a common absolute slope $k$.  This was done by minimizing
\begin{equation}
\chi^2 = \sum_i \left(\frac{d_{\mathrm{meas},i}\pm k R^\prime + c_j}{\sigma_i}\right)^2,
\end{equation}
where the terms in the sum that correspond to $\vecB$ down (up) used the + (-) sign and the intercept $c_1$ ($c_2$).  The fit yielded a crossing point at $R_\times = 3.50 \pm 0.81$, $d_\times = -0.66 \pm \dxUncert \times 10^{-26}$ \ecm, with $\chi^2/\nu$ = 651/542 and $k =(1.88\pm0.25)\times10^{-26}$ \ecm/ppm.  This fitted value of the slope was particularly influenced by one run (number 1900) that was taken at a large applied $\dBzdz$ (giving it significant leverage) and which departed from the fitted line by $2.7\sigma$: excluding that run would have changed the value of the fitted slope to $1.63\pm0.27 \times10^{-26}$ \ecm/ppm and increased the crossing-point EDM value by $0.22\times10^{-26}$ \ecm, but there appeared to be no {\it a priori} reason to do so.  The run was therefore included both in the initial 2006 analysis \cite{Baker_2006} and in the analysis that follows here.

According to \eqnabbr \ref{eqn:daf_Hgn}, the slope $k$ is expected to be inversely proportional to $\Delta h$.  We shall discuss the estimation of $\Delta h$ in \secabbr \ref{sec:deltah}.  The relationship between $R$ and $\dnf$ is also affected substantially by the phenomenon of gravitationally enhanced depolarization \cite{Harris_2014}, and this will be discussed in \secabbr \ref{sec:grav_depol} below. In addition, the slope $k$ can be altered by a few percent (although still remaining highly symmetric under $\vecB$\ reversal) by various other mechanisms including the UCNs' own false-EDM signal (a 2\% effect); a slight reduction in mean free path due to cavities and grooves in the electrodes as well as to the presence of 10$^{-3}$ torr of He gas to prevent sparks \cite{Pendlebury_2004, Lamoreaux_2005} (1\%); and a possible bias in the volume-averaged frequency measurement due to asymmetric surface relaxation of the Hg, e.g.\ if there is a preferential depolarization on one or other electrode (up to 5\%).

\subsection{Estimating $\Delta h$}
\label{sec:deltah}

As seen in \eqnabbr \ref{eqn:daf_Hgn}, the slope $k$ of the lines in \figabbr \ref{fig:edm_vs_R} is determined, to first order, by the height difference $\Delta h$ between the centers of mass of the mercury and neutron ensembles.  Initial estimates\cite{Baker_2006} yielded $\Delta h = 2.8 \pm 0.1$ mm.  This was based on a misinterpretation of the frequency response in a trap with a variable-height lid, arising from a lack of understanding at the time of the way in which gravitationally enhanced depolarization affects the measured neutron-to-mercury frequency ratio.  A detailed Monte Carlo simulation has now been carried out, taking into account the known properties of the neutron source, guide tubes, elbow, polarizer foil, and storage chamber; it reveals that the spectrum was considerably softer, i.e.\ of lower mean energy, than had originally been assumed.  

The simulated spectrum is also essentially consistent with a largely analytic calculation  \cite{Baker_2014}, which concluded that at the start of the storage period  the UCN density $\rho_\epsilon(z)$ at the level $z=0$ of the lower electrode could be roughly approximated by a softened Maxwell spectrum, 
\begin{equation}
\label{eqn:soft_maxwell}
\rho_\epsilon(0) = A\cdot \epsilon^{\frac{1}{2}}\exp\left(-\epsilon/V_f\right),
\end{equation}
where $V_f$ = 88 neV is the Fermi potential of the quartz insulator, above which energy $\rho$ drops to zero.  Henceforward, when we refer to the UCN energy, it is to be taken to mean the kinetic energy of the UCN at the level of the bottom electrode; or, equivalently, the total (kinetic plus potential) energy at any point in the bottle, where the potential energy is again referenced to the level of the bottom electrode. \eqnabbr \ref{eqn:soft_maxwell} estimates relative densities at the bottom electrode, rather than total numbers in the bottle: the latter would have a slightly different distribution, since higher-energy neutrons extend further in height.  In order to account for this, we note that for a mono-energetic group having energies between $\epsilon$ and $\epsilon+d\epsilon$, where the energy  $\epsilon$ is given in terms of the maximum height that the neutrons could reach if travelling vertically upwards,  
the available phase space dictates that the density variation with height $\rho_\epsilon(z)$ 
has the form \cite{Pendlebury_1994}
\begin{equation}
\label{eqn:rho_epsilon}
\rho _\epsilon (z)=\rho _\epsilon (0)\left[ {\frac{\epsilon-z}{\epsilon}} \right]^{1/2},
\end{equation}
and one must integrate this function over the bottle height (or, if less, the attainable height $\epsilon$) to establish the relative numbers of UCN of each energy trapped within the bottle. 

\begin{figure}[ht]
  \begin{center}    
\resizebox*{0.5\textwidth}{!}{\includegraphics {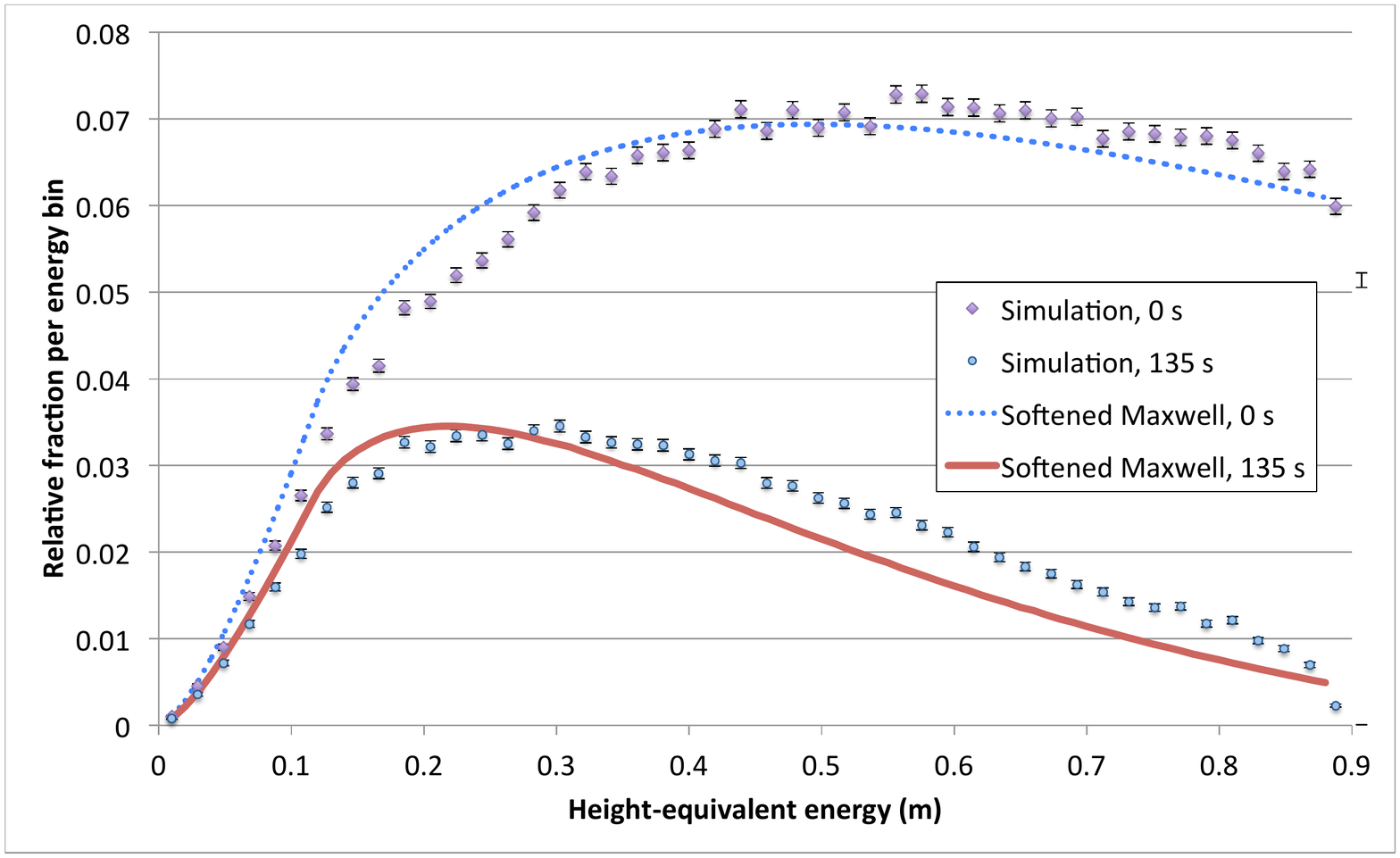}}
    \caption{
(Color online) Calculated energy spectrum of the stored UCN, both at the start ($t = 0$) of the storage period (blue dotted line: softened Maxwell spectrum; purple diamonds: simulation) and after $t =135$\,s (solid red line: softened-Maxwell spectrum; blue circles: simulation).  The total area under the simulated spectrum at $t = 0$ s is normalised to one; the softened-Maxwell spectra are scaled to match the simulated peak height at $t = 135$ s.
}
    \label{fig:spectrum}
  \end{center}
\end{figure}

Although our best estimate of the spectrum is that given by the simulation -- which will underlie the analysis that follows -- we have carried out a complete analysis for both spectra, in order to see whether there was any significant sensitivity of the final result upon the initial spectrum.   These two initial spectra are shown as the uppermost line and set of points  in \figabbr \ref{fig:spectrum}.

\begin{figure}[ht]
  \begin{center}    
\resizebox*{0.5\textwidth}{!}{\includegraphics {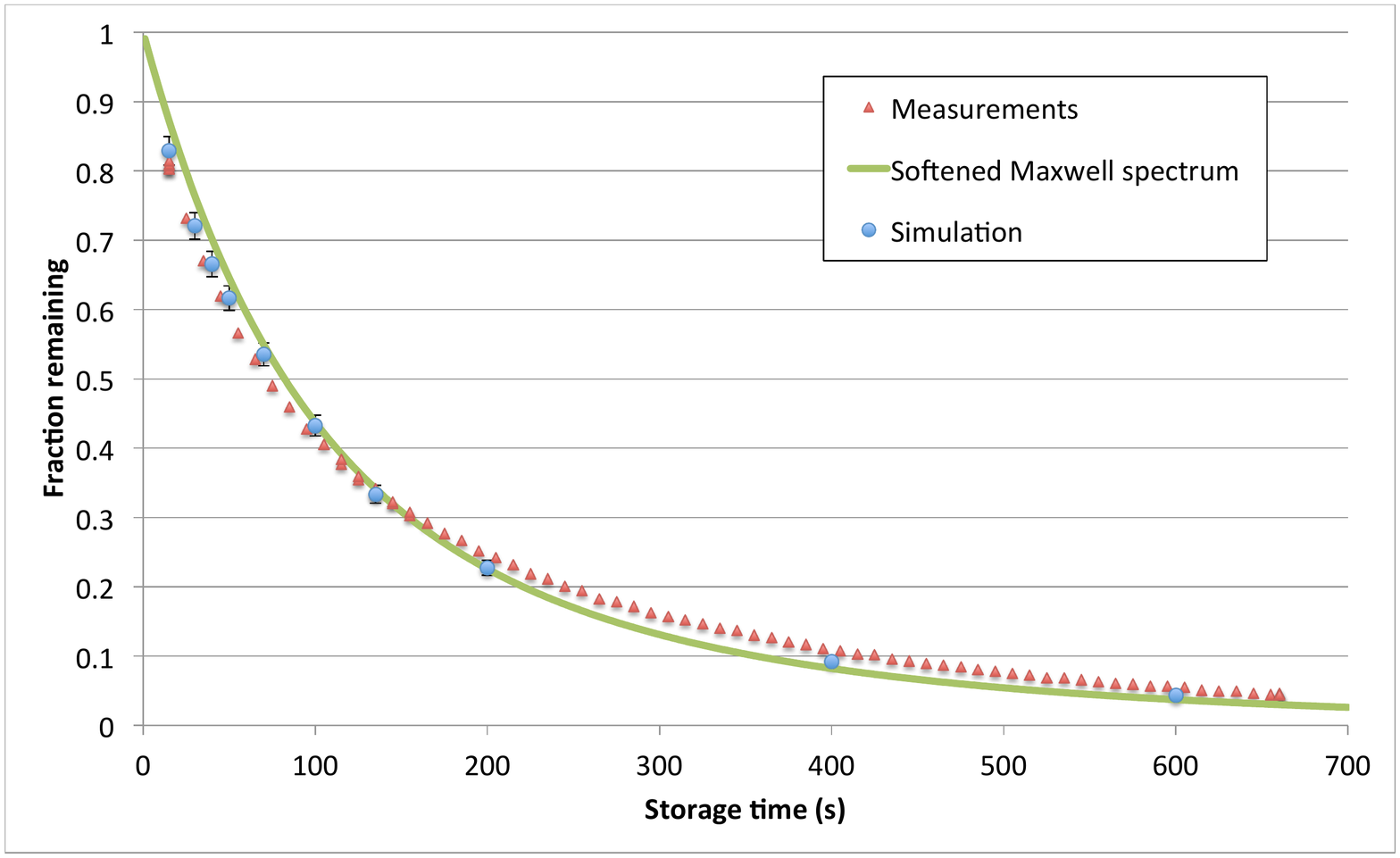}}
    \caption{
(Color online) Fraction of neutrons remaining in the bottle as a function of the time for which they are stored.  Red triangles represent measurements; blue circles and the solid green  curve represent retained fractions for the simulated and the softened-Maxwell spectra, respectively.
}
    \label{fig:storage_time_curve}
  \end{center}
\end{figure}

The spectrum softens during storage, since higher-energy UCN not only have a higher interaction rate with the walls; they also have a higher loss probability per bounce, as given by \cite{golub_UCN_book}
\begin{equation}
\bar{\mu}(\mathcal{E}) = 2f\left[ \frac{V}{\mathcal{E}}\sin^{-1}\left(\sqrt{\frac{\mathcal{E}}{V}}\right) - \sqrt{\frac{V}{\mathcal{E}}-1} \right],
\end{equation}
where $f = W/V$ is the ratio of the imaginary ($W$) to the real ($V$) part of the material's Fermi potential.  In fact, although $V$ is in general well known, $W$ is difficult to determine; and in any case, losses are likely to be dominated by hydrogen that has diffused into the containing surfaces.  This hydrogen can -- because of its extremely high incoherent scattering cross section -- substantially influence loss rates without significantly altering the surface potential $V$.  The value of $f$ used in the simulation was therefore adjusted until, at $f=4\times10^{-4}$, the loss rate (including $\beta$ decay) more or less matched the observed loss rate of neutrons in the bottle, as shown in \figabbr \ref{fig:storage_time_curve}.  This is broadly similar to the value $f=3\times10^{-4}$ obtained from an independent data-simulation comparison undertaken with the apparatus circa 2006 \cite{kuzniak08}.   It was assumed in these calculations (as in \cite{kuzniak08}) that $f$ would be similar for both the insulator and the electrodes.  The softened-Maxwell spectrum was propagated in a similar manner (although for this $f=3\times10^{-4}$ gave a better fit to the storage-time data, as shown also in \figabbr \ref{fig:storage_time_curve}).  The resulting two spectra after 135 s of storage, by which time the Ramsey sequence is complete, are shown along with the initial spectra in \figabbr \ref{fig:spectrum}.  The resulting final simulated spectrum is slightly firmer than the final softened-Maxwell spectrum.

Throughout this analysis the storage trap was modeled as a simple cylinder, with an additional cavity at the bottom centre,  4.0\,cm deep and 3.4\,cm in radius, within which is set the UCN entrance door to the trap.  



Given that the calculated $\Delta h$  after storage is \deltahAfterStorage\, mm, to first approximation the  anticipated slope $k$ of the lines of \figabbr \ref{fig:edm_vs_R} should decrease from $1.57\times10^{-26}$ $e$cm/ppm (in \cite{Baker_2006}) to  $\ExpectedSlope\times10^{-26}$ $e$cm/ppm, before taking gravitationally enhanced depolarization into account.

\subsection{Gravitationally enhanced depolarization}

\label{sec:grav_depol}

\begin{figure}[ht]
  \begin{center}    
\resizebox*{0.5\textwidth}{!}{\includegraphics {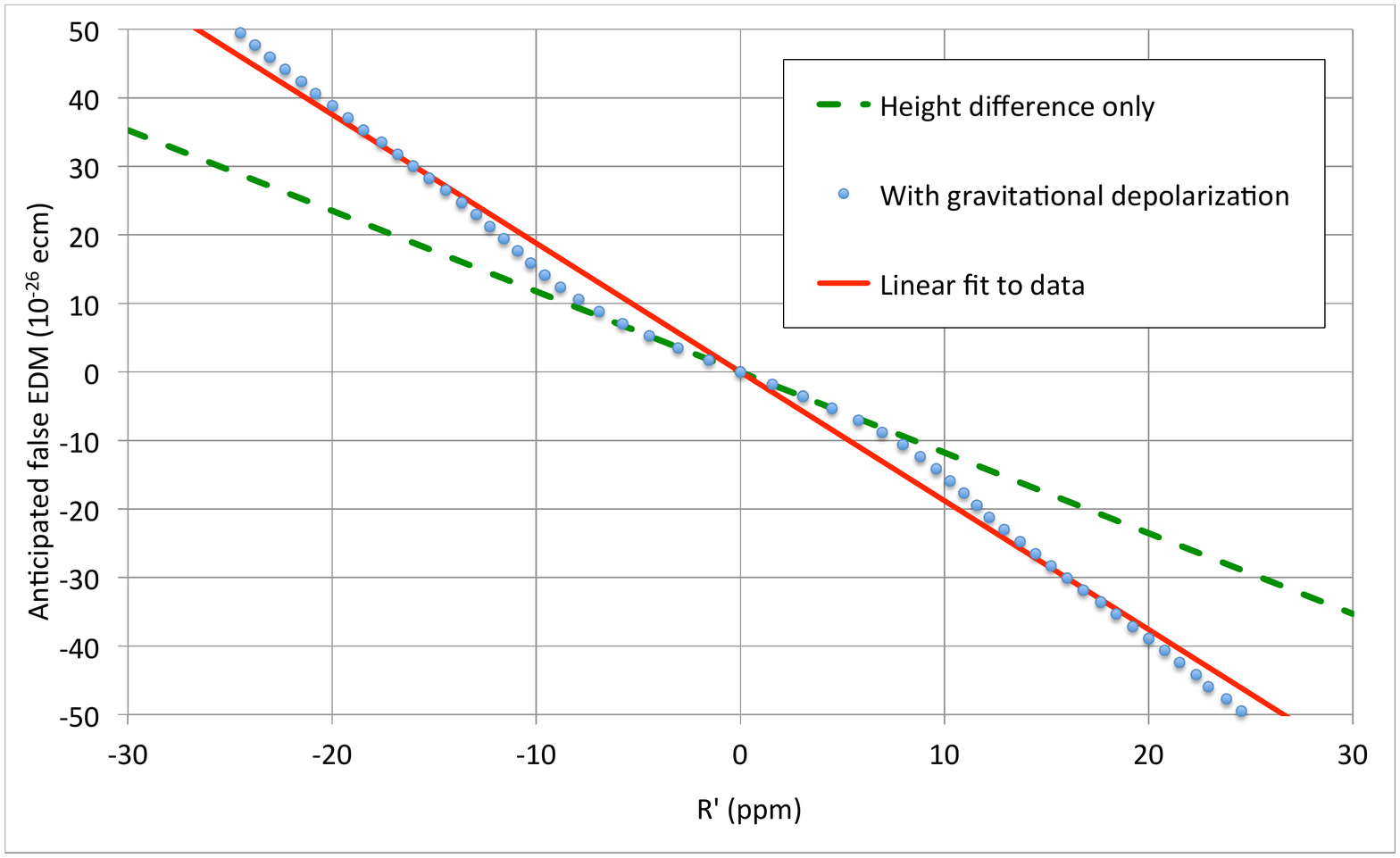}}
    \caption{
(Color online) Plot showing the anticipated false edm $d_\mathrm{n,Hg,f}$ as a function of $R^\prime$, for $\vecB$ upwards.  The dashed green line is the expectation from \eqnabbr \ref{eqn:daf_Hgn}, and the blue dotted line shows the revised expectation when gravitational depolarization is taken into account.  The red line is the linear best fit to the data, as shown in \figabbr \ref{fig:edm_vs_R}. 
}
    \label{fig:dnf_vs_R_grav_depol}
  \end{center}
\end{figure}

Under the influence of gravity, neutrons of different energies effectively sample different regions of the storage trap.  The (sometimes surprising) consequences of this effect have only recently been studied in detail \cite{Harris_2014, knecht09}, and have now been validated by comparison with data \cite{Afach_2015d, afach15b}.  This stratification results in a dephasing of the ensemble in addition to that arising from the intrinsic depolarization that naturally occurs within each energy bin.  Since lower-energy neutrons preferentially populate the bottom of the bottle, this dephasing is asymmetric, and results in a (strongly spectrum dependent) frequency shift.  Calculations in \cite{Harris_2014} were based upon a Maxwellian velocity distribution with a sharp 93\,cm height-equivalent energy cutoff.  The simulated spectrum shown in \figabbr \ref{fig:spectrum} is much softer than this, showing a clear peak at 20-30\,cm height-equivalent energy.  Under these conditions the gravitational-depolarization effect distorts the shape of the crossing lines, significantly enhancing the slope for measurements beyond a few\,ppm from the origin, as shown explicitly in \figabbr \ref{fig:dnf_vs_R_grav_depol}.  The solid red line is the best linear fit to the data, mirroring the lines shown in \figabbr \ref{fig:edm_vs_R}, whereas the dashed green line is the expectation from \eqnabbr \ref{eqn:daf_Hgn}.  The dotted blue line includes the effect of gravitationally enhanced depolarization; we designate its functional form as 
\begin{equation}
d_\mathrm{n,Hg,f} = \xi(R^\prime).
\end{equation}

The data of \figabbr \ref{fig:edm_vs_R} were fitted to 
\begin{equation}
d_\mathrm{meas} = d_\times\pm k^\prime \xi(R^\prime-R_\times),
\end{equation} 
thus allowing the crossing-point coordinates $R_\times, d_\times$ to vary, as well as allowing $\xi$ to be multiplied by a factor $k^\prime$ as a consistency check of the input spectrum, since the slope of the function is completely determined by the spectrum via $\Delta h$.   As well as crossing-point values of $R_\times = \RxValueFromThreeParamLineFit$\,ppm, $d_\times = (\dxValueFromThreeParamLineFit)  \times 10^{-26}$ \ecm, the fit yielded $k^\prime = \kprime \pm \kprimeUncert$; and, when it was repeated without inclusion of the aforementioned possibly anomalous datum of Run 1900, we found $k^\prime = \kprimeNoDubiousRun$.  These numbers are consistent with unity, giving confidence in the simulated spectrum (for comparison, the softened-Maxwell spectrum yielded $k^\prime = \kprimeSoftMax \pm \kprimeUncertSoftMax$), and so, going forward, we set $k^\prime = 1$.  We therefore fitted the data instead to 
\begin{equation}
d_\mathrm{meas} = d_\times\pm \xi(R^\prime-R_\times).
\end{equation} 
This yielded crossing-point values of $R_\times = \RxValueBeforeRBSCorr \pm \RxFitUncert$\,ppm and $d_\times = (\dxValue \pm \dxUncert) \times 10^{-26}$ \ecm.  $\chi^2/\nu$ had a value of 650/542 = 1.19, unchanged from the earlier fit.  We note that the change from three to two free parameters was accommodated almost entirely by shifting $R_\times$, with virtually no change in $d_\times$; we deduce that the crossing-point value of the measured EDM $d_\times$ is relatively insensitive to such changes in the slope, and therefore also to the detailed shape of the input spectrum. 

By eye it may appear that the straightforward linear fit to the data is so close to the predicted curve that it should make little difference to any analysis.  However, the central linear region of the predicted curve is now shallower by a factor  \CentralRegionSlope/1.88 = \SlopesRatio\  compared to the slope of the original linear fit, and any systematic shifts in $R$ will therefore have a correspondingly reduced effectiveness in producing systematic shifts in $d_\times$.

A number of assumptions were made while calculating the size of the predicted false-EDM effect -- for example, the gravitational-depolarization calculation here assumes only a uniform vertical $B$-field gradient; and no allowance has been made for the effect of grooves in the electrodes (which are necessary to locate the insulating trap walls).  Effects of intrinsic depolarization were explicitly included in this model, although perhaps not perfectly since (as discussed in \cite{Harris_2014}) it has a significant dependence upon the specularity of reflections within the trap.  However, its contribution to the frequency-shift effect is negligible \cite{Afach_2015d}; primarily, it simply smears the phases.   Additionally, it is worth noting that in principle  the effective UCN spectrum could be slightly influenced, leading to different slopes $k_\uparrow$, $k_\downarrow$ for the two $\vecB$ field directions,  if the spin relaxation were substantially different for the two configurations.  This extra relaxation, if it existed, would be expected to be UCN-velocity dependent, and so would change the effective UCN energy spectrum.  However, the observed polarisation product $\alpha$ (i.e. the visibility of the Ramsey fringes) is identical at the 2\% level for the two $B$-field polarities.  This suggests \cite{Afach_2015d} that the average UCN energy, and therefore $\Delta h$ (which is approximately inversely proportional to the energy) should be the same to better than a few percent between the two polarities.  It is difficult to conceive of a mechanism that would change the slope more significantly than this upon field reversal.  

We have one further small correction to make at this point.  Since the Ramsey-resonance measurements were carried out with an oscillating rather than a rotating transverse RF field, the measured frequencies are subject to a Ramsey-Bloch-Siegert (RBS) shift \cite{Ramsey_1955,Bloch_1940,may_phd} given by 
\begin{equation}
\Delta R_{RBS} = \frac{\pi}{4\omega_0^2tT\left(1+\frac{8t}{\pi T}\right)},
\end{equation}
where $\omega_0$ is the resonant frequency (182 rad/s), $t$ is the period (2 s) for which each of the RF pulses is applied, and $T$ is the (130 s) period of free precession between the RF pulses.  This amounts to a mere 0.09\,ppm.

Overall, therefore, we regard the match between the measured and expected curves to be in reasonable agreement.  Going forward, we use the (RBS-corrected) value $R_\times = \RxValue \pm \RxFitUncert$\,ppm, along with  $d_\times = (\dxValue \pm \dxUncert) \times 10^{-26}$ \ecm, as input to our subsequent analysis.  


\subsection{Statistical sensitivity}

For the majority of the data-taking runs, the upper electrode was held at voltages of circa $\pm$ 80 kV, giving an applied electric field of $E$ =  7 kV/cm.  (Voltages as high as 130 kV were sustainable towards the end, whereas in early runs it was not possible to go much above 60 kV.)  The average polarization was $\alpha = 0.58$, and the Ramsey coherence time was $T$ = 130 s.   Removing from consideration the cycles measured at $E = 0$ kV/cm, \eqnabbr \ref{eqn:edm_uncert} yields an anticipated sensitivity of $ 1.34 \times 10^{-26}$ \ecm.  Our achieved sensitivity is a few percent larger than this, for a number of reasons: for example, field changes meant that not all cycles were taken at the intended points on the Ramsey curve; some cycles had frequency errors enlarged by relatively rapid depolarization of the mercury; and any changes in field gradient resulting in changes in $R^\prime$ during the run (which at the $\sim$ 0.2\,ppm level or below would not be detectable) could also add noise at the percent level. 


\section{Modifications to the false EDM signal}
\label{sec:false_edm_mods}
 
As indicated above, there are some processes that can displace the crossing point, and thus interfere with this technique of removal of false-EDM effects  -- essentially any process that changes $R_a$ and/or $\dnf$ without conforming to the ratio between the two given by \eqnabbr \ref{eqn:daf_Hgn}, and where, in addition, the changes differ with the direction of $\vecB$. These processes may broadly be divided into two categories: those that shift frequencies, and hence move the lines of \figabbr \ref{fig:edm_vs_R} horizontally,  and those that move the lines vertically by altering the gradient-induced false EDM signal relative to the expectation from $\left<\dBzdz\right>$. In this section we summarize these effects, before describing in more detail in \secabbr \ref{sec:aux_meas} the additional diagnostic measurements that were carried out first to characterize and then to compensate  for them.  Other systematic effects, which contribute to the overall uncertainty but that do not introduce a bias to the data, are considered in \secabbr \ref{sec:other_syst}.
   
\subsection{Uncertainty in $\gamma_\mathrm{n}/\gamma_\mathrm{Hg}$}
  An offset in $\gamma_\mathrm{n} / \gamma_\mathrm{Hg}$ shifts the two lines of \figabbr \ref{fig:edm_vs_R}\ sideways in the same direction by the same amount, leaving $d_\mathrm{meas}$ unaffected.  This therefore has no direct effect on the final result, although it can influence slightly the overall fit to the data.  

\subsection{Horizontal quadrupole fields}
      Fields with finite $\partial B_x/\partial y$ and/or $\partial B_y/\partial x$ but with ($\partial B_x/\partial x + \partial B_y/\partial y) = 0= -\dBzdz$ cause $R_a$ to increase quadratically \cite{Pendlebury_2004} without contributing to $\dnf$.  This arises because the neutron spins follow the total field direction adiabatically, with a correspondingly increased precession frequency, whereas the (nonadiabatic) Hg atoms average out the transverse components and are generally sensitive only to the $B_z$ component.  We consider here a quadrupole field aligned with $z$, with $B_x = qy, B_y = qx$.  This gives rise to a shift in $R_a$ of 
\begin{equation}
\delta R_Q = \frac{q^2 r^2}{4B_0^2},
\end{equation}
where, as before, $r$ is the radius of the trap.
  
As discussed in \cite{Baker_2014}, the region in the vicinity of the EDM measurement volume was scanned with a fluxgate magnetometer.  This revealed the presence of such horizontal field components, quadrupolar in form and a few\,nT in magnitude throughout the 12\,cm height of the measurement bottle.  The magnitude of these quadrupole fields also depends slightly upon the orientation (upwards or downwards) of the  $\vecB$ holding field.  This is to be expected, since the $\vecB$ return flux passes through the innermost mu-metal shield: indeed, about 40\% of the $\vecB$ field arises not from the $\vecB$ coil directly but from the magnetization of the shields. The geometry of the 1.15 m diameter $\mu$-metal shield around the $\vecB$ coil has irregularities on the scale of a few mm, so it seems reasonable to expect that aberrations in the $\vecB$ field near the outer edges of the trap should be of the order of a few nT.  Unlike the remanent fields arising from any permanent magnetization of the shields themselves, any $B_{0xy}$ fields will reverse when $\vecB$ is reversed.  Contributions from these two different sources may therefore add in one field direction and subtract in the other, leading to a differential shift of all $R_a$ values,  and thus of the two lines, thereby changing the EDM value $d_\times$ at the crossing point of \figabbr \ref{fig:edm_vs_R}.

\begin{figure}[ht]
  \begin{center}    
\resizebox*{0.5\textwidth}{!}{\includegraphics {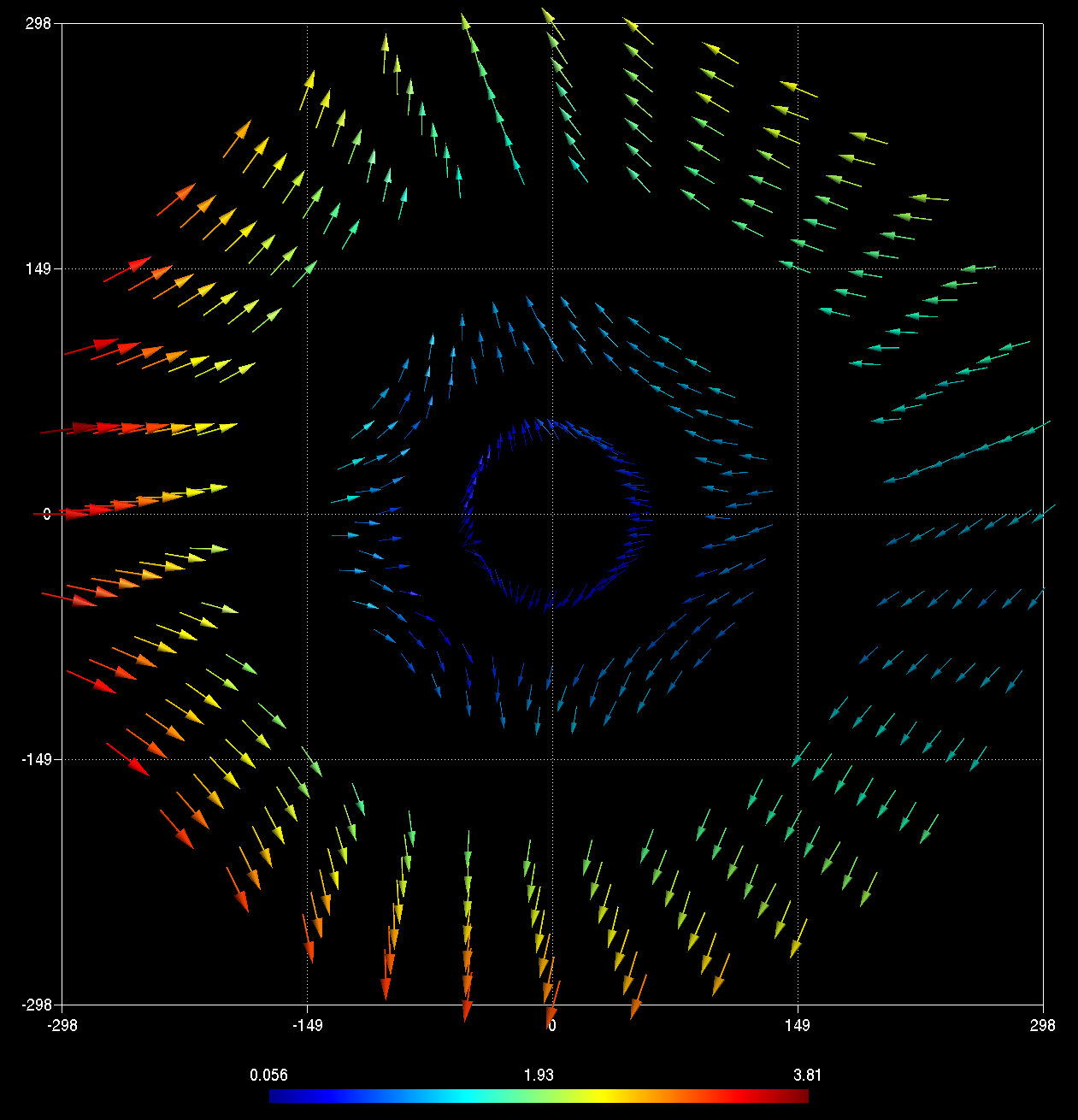}}
    \caption{
(Color online) Results of a detailed scan of the magnetic field inside the shields, taken some years after the EDM measurements described in this paper.  By this time the shields had been disassembled, moved to PSI and reassembled.  The quadrupole-field pattern is clear, and is obviously associated with the shields themselves.  The scales on the $x$ and $y$ axes represent the horizontal coordinates, in mm.  The magnitude of the field is represented both by the lengths of the arrows and, equivalently, by their colors, as shown in the scale (in nT) at the bottom.
}
    \label{fig:psi_scan}
  \end{center}
\end{figure}

Further evidence for the existence of such quadrupole fields came some years after the EDM measurements described in this paper, when the shields were disassembled, shipped to the Paul Scherrer Institut (PSI) and rebuilt.  The results of a detailed fluxgate scan taken in 2010 within the rebuilt shields is shown in \figabbr \ref{fig:psi_scan}, and here a quadrupole-type field pattern can be seen clearly.  A detailed harmonic analysis of the field shape yielded an average $q$ value of -1.29 (3.24) nT/m for $\vecB$ up (down), which would imply corresponding quadrupole shifts of $\delta R_Q$ = 0.02 (0.14)\,ppm for the two respective $\vecB$ directions.   We do not include these figures in our analysis at this point, but we will return to the issue briefly during our conclusions.

\subsection{Rotation of the Earth}   
 The rotation of the Earth shifts all of the frequency ratio measurements $R_a$ to lower values by 1.33\,ppm when the $\vecB$ field is upwards, and to higher values by 1.33\,ppm when the $\vecB$ field is downwards \cite{Lamoreaux_2007,Baker_2007}.  This therefore acts in a manner very similar to that of the horizontal-quadrupole shifts.

\subsection{Localized losses}
  $B$-field averaging in the trap is affected by localized loss of UCN and Hg particles, and by polarization loss in the presence of the $10^{-3}$ fractional $\vecB$ inhomogeneities, which may change with $\vecB$\ direction. However, we estimate that the resulting $R_a$ shifts are $< 0.1$\,ppm and $< 0.01$\,ppm for the UCN and Hg respectively.  Furthermore, they will be indistinguishable from the quadrupole shifts.  We do not consider them further.

\subsection{Magnetic dipoles}
\label{sec:dip_enhance}
   As demonstrated by \eqnabbr \ref{eqn:nonadiabatic_arb_field_false_edm}, the field of a permanent magnetic dipole (PMD) close to the trap results in a non-uniform $\dBzdz$, and therefore induces a frequency-ratio shift $\delta R_\mathrm{dip}$, moving the lines of \figabbr \ref{fig:edm_vs_R} horizontally by equal and opposite amounts for the two directions of $\vecB$.  It also adds to $\dnf$ a false EDM $d_\mathrm{dip}$ \cite{Harris_2006}, thus shifting the lines vertically by this amount.


Our fluxgate magnetometer surveys of the trap cannot rule out PMD fields of less than 1\,nT at 2\,cm from the inner surface. Large areas of the trap are SiO$_2$ or Al, backed by large voids, and do not come under suspicion; but the mercury and UCN doors involve a heterogeneous collection of small parts close to the trap. 

Based upon our simulations \cite{Pendlebury_2004,Harris_2006} (since confirmed to be perfectly consistent with the analytic prediction of \cite{Pignol_2012}), we included in our 2006 analysis a $d_\mathrm{dip}$ uncertainty of $\pm 6.0 \times10^{-27}$ \ecm\ to allow for an undetected 1\,nT PMD at the mercury door.   

 In the case of the UCN door we have better diagnostics. As discussed above, this door sits at the bottom of a small cavity, 4.0\,cm deep and 6.8\,cm in diameter,  at the center of the lower electrode. As detailed in  \secabbr \ref{sec:aux_meas}, measurements were made of the neutron and mercury response in a storage trap with a ceiling of variable height, and the results provided clear evidence  for the existence of both dipole and quadrupole fields.  This conclusion was reinforced by measurements undertaken some years later when, as part of the process of characterizing the system during its removal to PSI, scans were made of the components.  A crank-and-sliding-plate assembly that converted rotation of the driving shaft into the linear motion of the neutron door was revealed to be slightly magnetic, with summary records showing that a sensor several\,cm away registered a field of 2.5 nT.  A disk that constituted the lower part of the door itself also yielded a total field of about 100 pT at a distance of $\sim 13$\, cm from the sensor.  Various other parts, designated ``screws'', ``piston'', ``Al frame'' and so on, were tablulated as having associated fields of a few pT to a few tens of pT.     Detailed records of orientations and positions were not kept, since the primary goal at that time was merely to determine whether or not the components were magnetic.  It is therefore not possible at this point to specify the dipole position and strength precisely from the information now available, so as in the 2006 analysis we rely upon the contemporary measurements and regard the later scans merely as qualitative supporting evidence, but it seems impossible to avoid the conclusion that there were dipole-like fields of $\sim$ nT strength in the vicinity of the UCN door.  The picture is also consistent with measurements undertaken (after the EDM measurements, but prior to moving the apparatus to PSI) with a fluxgate magnetometer located in the vertical neutron guide immediately underneath the neutron door.  This registered a slight change (circa 1 nT) in the ambient magnetic field when the door mechanism was operated.

The dipole-field analysis that we have carried out assumes a vertically orientated dipole centered on the axis.  There is no particuar reason to assume that any PMD would have this particular orientation or precise radial position.  However, we note that the associated frequency shifts arise via $\langle \partial B_z/\partial z \rangle$, which would be much lower in the case of a horizontal orientation (not least due to cancellation of the respective field components  from opposite ends of the dipole).  Furthermore, the analysis of \cite{Pignol_2012} (and in particular \figabbr 2 therein) shows both that a horizontal alignment (for a dipole within a few cm of the axis) yields a much smaller EDM contribution than does a vertical alignment, and also that the EDM contribution (of the vertically aligned dipole, again within a few cm of the axis) has very little sensitivity to the radial position.  We therefore conclude that our model accounts in a  reasonable manner for the most important contributions.

The post-hoc magnetic-field scans at PSI revealed no magnetic contamination in the vicinity of the mercury door, although a re-scan in 2011 showed a small contaminant outside the storage volume, at a radius of 31 cm, within the threaded hole connecting the ground corona ring to the lower electrode.  If present throughout our measurements, this could have contributed a false EDM of up to $\pm 4.0 \times10^{-28}$ \ecm\ (according to the Pignol-Roccia analysis).  In order to be conservative, though, we retain the full $\pm 6.0 \times10^{-27}$ \ecm\  contribution to the systematic uncertainty, in order to accommodate both the possible undetected presence of a (further) PMD as well as the fact that we have not modelled the actual door dipole perfectly.


\subsection{Mercury light shift}
\label{sec:lightshift}

The presence of the mercury reading light can, via the RBS mechanism \cite{Ramsey_1955,Bloch_1940}, shift the resonant frequency of the mercury atoms\cite{Cohen-Tannoudji_1972, corney}. Such shifts are produced by any small component, parallel to $\vecB$, of the $^{204}$Hg probe light beam passing through the precessing $^{199}$Hg atoms. This component, and the consequent $R_a$ shift, reverse sign on reversal of $\vecB$. False EDM signals can then arise in two distinct ways:
\begin{itemize}
 \item{} Direct: If any changes in intensity are correlated with the electric field direction, the resultant frequency shift would mimic an EDM.  
 \item{} Indirect: Even if the intensity of the light is completely constant, the light shifts move the data points of the two lines in  \figabbr \ref{fig:edm_vs_R} in opposite (horizontal) directions, and thus contribute a vertical offset to the crossing point.  
\end{itemize}

A slight dependence of $R_a$ on the incident light intensity was indeed found during the 2006 analysis, the magnitude $\sim 0.2$\,ppm being in agreement with theory. (Similarly small shifts were found in a recent measurement of the ratio of neutron to $^{199}$Hg gyromagnetic ratios, using the upgraded nEDM apparatus at PSI with similar Hg discharge bulbs but with a new storage chamber \cite{Afach_2014a}.)  A correction to $d_\mathrm{meas}$ was made on a run-by-run basis, leading to an overall correction of $(3.5 \pm 0.8) \times 10^{-27}$ \ecm.  For this analysis we have simply applied exactly the same set of (almost imperceptible) shifts to the data, prior to the crossing-lines fitting procedure.   Since the dependence should thereby have been removed, we expect no remaining net bias from this source.  Details of the light-shift analysis are discussed extensively in \cite{Baker_2014}, and will not be addressed further here.

\section{Auxiliary measurements}
\label{sec:aux_meas}

  Separate $\nu_\mathrm{n}$, $\nu_\mathrm{Hg}$  and $R_a$ measurements (without $E$ fields) were made in an auxiliary trap with a roof that could be raised or lowered to change the height $H$. Critically, the trap was built on the same lower electrode and door mechanism that were used for EDM data taking.  Full details of the apparatus and of the measurements made are available in \cite{richardson_phd}.

The trap had a smaller radius than the main data-taking bottle, 18.5\,cm rather than 23.5\,cm, and because the floor of the trap consisted of an aluminum plate that rested on the lower electrode, the door-cavity depth became 6.0 rather than 4.0\,cm. These differences would have  reduced the quadrupole shifts by a factor $Q \sim 0.59$.  (Note that, although the quadrupole shifts are reduced in this way, the shifts due to Earth's rotation are not.) The roof and floor were coated with deuterated polystyrene (DPS) rather than the diamond-like carbon used for the main data-taking trap, but the DPS tended to flake away, leaving some bare aluminum exposed: the resulting UCN spectrum would therefore have had a slightly lower end-point cutoff than in the data-taking trap, although after some tens of seconds of storage this would make little difference.  Measurements were taken at storage times of several tens of seconds' duration rather than the full 135\,s, because with the reduced-height bottle the loss rates were much higher.  Spectrum calculations were carried out exactly as described above for the data-taking bottle, assuming the same initial spectrum and per-collision loss rates.  Appropriate Ramsey-Bloch-Siegert frequency-shift corrections \cite{Ramsey_1955,Bloch_1940}, typically amounting to ~0.1-0.2\,ppm,  were applied to each measurement.

 Assuming $B_z(z) = b_0 + b_1z + b_2z^2$, one can show that $\langle\dBzdz\rangle_V$ = 0 for a trap height $H$ when $\nu_\mathrm{Hg}$ is the same for roof settings at $H/2$ and $H$.  This situation was approximated by adjusting the trim coils until the same $\nu_\mathrm{Hg}$ was obtained for trap heights of 109 and 59 mm.  With this field established,  $R^\prime$ was measured for trap heights $H$ of 34, 59, 84 and 109 mm for $\vecB$ down.  The procedure was repeated for $\vecB$ up, although in this case measurements were only made for trap heights of 59 and 109 mm.   We denote the values of $R^\prime$ in this ``gradient-compensated'' environment as $R^\prime_\mathrm{gc}(H)$.  Naturally, it was not possible in reality to trim the gradients perfectly, but studying the Hg response enabled appropriate corrections to be made.  For this analysis we have restricted ourselves to the eleven measurement runs for which such gradient corrections to $R^\prime$ amount to less than 0.5\,ppm.
   
\begin{figure}[ht]
  \begin{center}
    \resizebox*{0.5\textwidth}{!}{\includegraphics
    {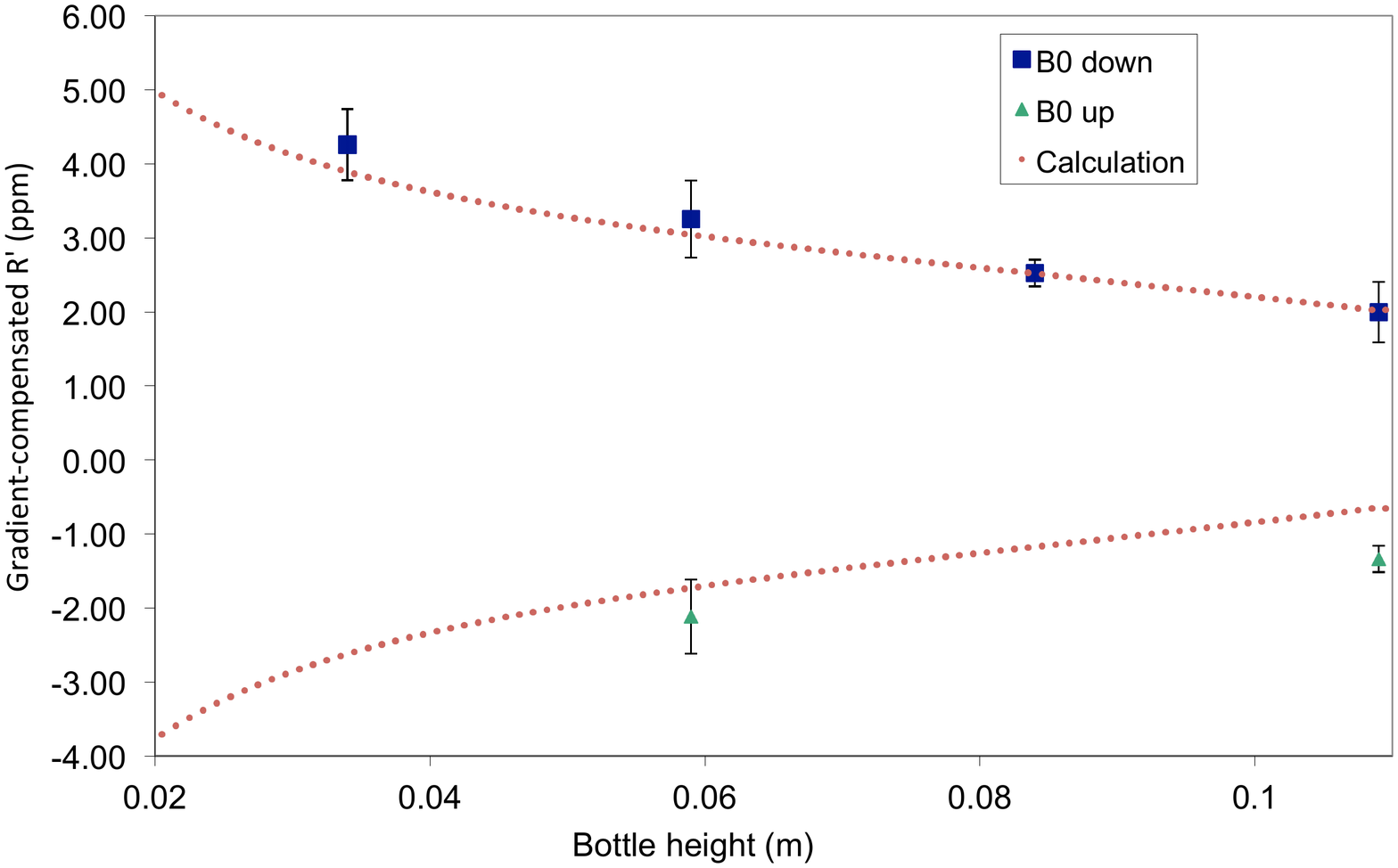}}
    \caption{(Color online)  Values of $R^\prime$ (in\,ppm) for a ``gradient-compensated'' field (see text) for each $\vecB$-field polarity, as a function of the height of the storage volume.  Where more than one datum was taken for a particular configuration, the point shown represents the average.  
The curved dotted lines represent an {\em approximation} to the fitted function: they are based upon the spectrum calculated for 75\,s of storage in a bottle of height 84\,mm, whereas the fit uses spectra specific to each datum. }
    \label{fig:aux_trap}
  \end{center}
\end{figure}

   The resulting values of $R^\prime_\mathrm{gc}(H)$ are listed in \tababbr \tabnumRvsH. Their averages for each height and magnetic field configuration are shown in \figabbr \ref{fig:aux_trap}.  The original intention of this series of measurements had been to determine a value for the ratio of gyromagnetic ratios $\gamma_\mathrm{n}/\gamma_\mathrm{Hg}$. However, the strong variation with height, and in particular the separation of $\vecB$ up and down values as the height was reduced, led to the conclusion that there was a dipole field of strength $\sim 1$\,nT penetrating into the door cavity.   Such a field could in principle have been produced by a pointlike contaminant more or less at the limit of detectability by a fluxgate magnetometer, such as the aforementioned door-actuator crank.  It could also in principle have arisen from a more dispersed (and therefore undetectable) source such as, for example, a slight residual magnetization within the BeCu door plate or from a thermoelectric current as the door mechanism operates {\it in vacuo}.   


\begin{table}
\label{tab:R_vs_H}
	\centering
		\begin{tabular}{|c|c|c|c|c|}
		\hline
		$\vecB$ dir.		& $H$ (mm) 	& $T_\mathrm{meas}$ (s) &$R^\prime_\mathrm{gc}$ & $\sigma_R$  \\    \hline
		$\downarrow$ 		& 34  	& 30	& 4.80 & 0.73 \\ \hline
		$\downarrow$ 		& 34  	& 50	& 3.83 & 0.64 \\ \hline
		$\downarrow$ 		& 59	  	& 40	& 4.73 & 1.17 \\ \hline
		$\downarrow$ 		& 59	  	& 70	& 2.74 & 1.03 \\ \hline
		$\downarrow$ 		& 59	  	& 70	& 2.96 & 0.70 \\ \hline
		$\downarrow$		& 84		& 70	& 2.45 & 0.31  \\ \hline
		$\downarrow$		& 84		& 100	& 2.56 & 0.22  \\ \hline
		$\downarrow$		& 109	& 70	&2.42 & 0.59  \\ \hline
		$\downarrow$		& 109	& 70	&1.60 & 0.57  \\ \hline
		$\uparrow$		& 59		& 70	& -2.12 & 0.50  \\ \hline
		$\uparrow$		& 109	& 100	& -1.34 & 0.18  \\ \hline
		\end{tabular}
		\caption{Values of $R^\prime_\mathrm{gc}$, in\,ppm, as a function of the direction of the magnetic field $\vecB$, the height $H$ of the gradient-compensated auxiliary trap, and the Ramsey measurement time $T_\mathrm{meas}$.   }
\end{table}

\section{Global fit}
\label{sec:globalfit}

In this \secabbr we detail the manner in which each contributing component was modeled and incorporated into the fit, before moving on to discuss the outcome of the fit and the consequent corrections that must be applied to the crossing-point EDM measurement $d_x$. 

\subsection{Dipole field}

   A simple model incorporating a dipole of moment $p$ situated a distance $z_\mathrm{dip}$ below the surface of the door was found to fit the form of the auxiliary-trap data extremely well.  The average $z$ component $\left< B_z \right>$ of such a field over a circle of radius $r$ a distance $z$ above the dipole is
   \begin{equation}
   \label{eqn:dipole_Bz_avg}
   \left< B_z \right> = \left(\frac{p}{4\pi}\right)\frac{2}{\left(z^2+{r}^2\right)^{3/2}}.
   \end{equation}
By dividing the trap volume numerically into 1 mm thick discs, the additional contribution to $B_z$ from the dipole field and from an additional uniform applied gradient could be calculated as a function of the height $z$.  Assuming a spectral distribution of UCN as given in \secabbr \ref{sec:deltah} (in contrast to the 2006 analysis, which assumed a Maxwellian velocity distribution up to the quartz Fermi-potential cutoff), the relative height distribution of the neutrons in each energy bin was calculated.   The contributions to the mercury and UCN frequencies, and hence the expected ratio $R_a$, could then be calculated as a function of the height $H$ of the trap roof.  Each iteration of the fit used its selected values for the strength and position of the dipole, for $R_\gamma$, and for the quadrupole shifts.  An appropriate uniform gradient was then applied so as to yield the same calculated mercury response at $H = 59 $ mm and 109 mm, thus matching the experimental configuration of the auxiliary bottle.  The $R_a$ values for each bottle height $H$ were then calculated and compared with the data before proceeding to the subsequent iteration of the fit.  We designate the shifts in $R_a$  due to the gradient-compensated dipole field as $\delta R_\mathrm{dip}(H)$.

\subsection{Quadrupole fields}

The quadrupole-field shifts $\delta R_{Q\uparrow}$, $\delta R_{Q\downarrow}$ (where the subscript arrows indicate the respective $\vecB$ directions) are assumed to be independent of height (in reasonable agreement with the fluxgate measurements), and to be proportional to a scale factor $Q$ that depends in principle only upon the (volume-averaged) square of the trap radius, thus yielding $Q = 0.59$ for the auxiliary trap relative to the main data-taking trap as discussed in \secabbr \ref{sec:aux_meas} above. 

As noted above, these quadrupole-type fields almost certainly arise from the $\mu$-metal shields.  It is possible in principle that the dependence upon radius may not be as expected, particularly towards the outside of the data-taking bottle, which extends to about 40\% of the (58\,cm) radius of the inner mu-metal shield.    As a test, we allowed the $Q$ values for the auxiliary trap to be reduced by a further factor $f_Q$, which was introduced as a free parameter in the fit.  Ultimately this made no difference, as the fitted value of $f_Q = 0.95$ was close enough to unity that the final EDM limit was unaltered.  We therefore omitted this from the final analysis, i.e.\ we set $f_Q = 1$.

The quadrupole-field contributions $\delta R_Q$ to the $R^\prime$ values within the fit were therefore $Q \cdot  \delta R_{Q\uparrow,\downarrow}$ for the auxiliary bottle and $\delta R_{Q\uparrow,\downarrow}$ for the main data-taking bottle.


\subsection{Earth's rotation}
\label{sec:fit_earthrot_and_quad}
    As noted above, the rotation of the Earth shifts all of the frequency ratio measurements $R_a$ to lower values by $\delta R_E = 1.33$\,ppm when the $\vecB$ field is upwards, and to higher values by the same amount when the $\vecB$ field is downwards.  For either the auxiliary or the data-taking trap in isolation this effect would be indistinguishable from the differential quadrupole shifts, and in the 2006 analysis it was not taken into account separately.  However, when extrapolating from one trap to the other a correction has to be made since the Earth's rotation shift is independent of the trap radius and therefore of $Q$.  The shifts $\delta R_E$ are therefore built explicitly into the fit in this analysis. 
 
\subsection{Literature $\gamma$ ratios}
With no applied gradient and with any dipole and quadrupole contributions appropriately accounted for,  the true value $R_\gamma$ of $R^\prime$ should be recovered. The average of the two independent literature values of $|\gamma_\mathrm{n}/\gamma_\mathrm{Hg}|$, namely  3.8424574(30)  from \cite{Afach_2014a} and 3.8424560(66) derived from \cite{cagnac61} and \cite{Greene_1979}, is 3.8424572(27).  This was used as an input estimator of $R_\gamma$.  It is equivalent to $R^\prime = 0.31\pm 0.71$\,ppm.

\subsection{Crossing point}
\label{sec:xing_pt}

By symmetry, the average quadrupole shift $\delta R_Q = (\delta R_{Q\uparrow}+\delta R_{Q\downarrow})/2$ away from the true value of $R_\gamma$ should correspond to the crossing point $R_\times$ of the lines of \figabbr\ \ref{fig:edm_vs_R}: 
\begin{equation}
\label{eqn:Rx}
 R_\times = R_\gamma + \delta R_Q.
\end{equation}

It is convenient at this point to introduce the parameter
\begin{equation}
\label{eqn:RxQ}
 \delta R_{QQ} = \left(\delta R_{Q\uparrow}-\delta R_{Q\downarrow}\right),
\end{equation}
which represents the quadrupole-field splitting.

We note that, as shields were opened, closed and demagnetized many times throughout the four years of data taking, the profile of the field could in principle have changed during this time -- indeed, the same trim-coil settings could sometimes yield variations in $R_a$ values of $\pm 3$\,ppm.  However, since the overall $\chi^2_\nu$ of 1.2 for the fit to the lines of \figabbr \ref{fig:edm_vs_R} was not excessively large, such changes were clearly dominated by alterations in the vertical gradient, which would result in movement along the lines rather than displacement of data points from the lines.  This result is not surprising -- the largest holes in the magnetic shields were on top and underneath, and this would have provided the primary access route for ingress of external fields.  

We account for these random variations by increasing the uncertainty of the datum for the crossing point $R_\times$ by a factor $\sqrt{\chi^2_\nu} = 1.09$ in the global fit.

\subsection{Fitting procedure and outcome}

The rather simple underlying model of a dipole and a quadrupole field yields a fairly complex functional shape that can only be evaluated numerically.  The (five) parameters that were allowed to vary were  the dipole strength and $z$ position, $R_\gamma$, the quadrupole shift $\delta R_{\downarrow}$, and $\delta R_{QQ}$.  The (thirteen) input data points were (a) the literature value of $R_\gamma$, (b) the crossing point $R_\times$ from the earlier fit to the crossing lines, and (c) the eleven auxiliary-bottle measurements $R^\prime_\mathrm{gc}$ of $R^\prime$  in the gradient-compensated fields at various bottle heights $H$ and storage times $t_s$:

\begin{equation}
\label{eqn:R_aux}
R^\prime_\mathrm{gc} = R_\gamma + Q\delta R_{Q\uparrow,\downarrow} + \delta R_\mathrm{dip\uparrow,\downarrow}(H, t_s) + \delta R_E.
\end{equation}

The fit used the Levenberg-Marquardt algorithm.  The resulting best-fit function is represented, together with the data, in \figabbr \ref{fig:aux_trap}.  For clarity, the data in that plot are reduced to six points by averaging over each height and $\vecB$ up/down configuration, and the plotted function was generated with the spectrum appropriate to a 75\,s storage time in an 8.4\,cm high bottle.  

The fit yielded output values of $R_\gamma = \RgammaFit \pm \RgammaFitUncert$\,ppm, quadrupole-field shifts of $\delta R_{Q\downarrow} = \RqDown$\,ppm and $\delta R_{Q\uparrow} = \RqUp$\,ppm, 
and a small dipole of moment $m = \MagDipMom \times 10^{-7}$ Am$^2$ situated \MagDipDist\,cm below the surface of the door -- quite consistent with its being a part of the door mechanism.  Such a dipole would contribute a field \DipFieldAtDoor\,nT in magnitude 2\,cm above the surface of the door and 
\DipFieldAtFloor\,nT at the level of the electrode surface, and as such would be at or below the limit of detectability.  In fact, the fit was fairly insensitive to the dipole strength and position, with different starting values of the parameters yielding a range of dipole strengths  from 2-8 $ \times 10^{-7}$ Am$^2$ correlated with a range of positions from 0.8 to 2.2 cm below the door.  $R_\gamma$, $R_\times$ and $\delta R_{QQ}$, on the other hand, were robust.  

We note in passing that the value of $R_\times = \RxGlobalFit \pm \RxGlobalFitUncert$\,ppm that emerges is significantly shifted (approximately 1$\sigma$) from the input value of $\RxValue \pm 
\RxFitUncertChisqCorr$\,ppm.  However, since $d_\times$ is essentially independent of such small changes in $R_\times$, as discussed above, we do not pursue this line of enquiry further -- what we are interested in is shifts {\em relative} to $R_\times$.

  The overall $\chi^2$ of the global fit was 8.9 for the eight degrees of freedom.  The quality of the fit is therefore rather good, particularly bearing in mind the complexity of the fitting function, and suggests that the model is both sound and complete -- there is no evidence of contamination from fields of a different form.  


\subsection{Field-shift corrections}

In the absence of the dipole field, we would expect the $R^\prime$ values at which the vertical magnetic-field gradient within the data-taking bottle is zero to be determined by the respective quadrupole-field offsets for each of the two $\vecB$ directions, together with the shifts due to Earth's rotation:
\begin{equation}
\label{eqn:Rprime0}
R^\prime_{{0}\uparrow,\downarrow} = R_\gamma + \delta R_{Q\uparrow,\downarrow} \pm \delta R_E,
\end{equation}
where the upper (lower) sign corresponds to $\vecB$ down (up).  It would therefore be appropriate under those circumstances to determine the EDM $d_0$ at either of these points rather than at the crossing point ($R_\times, d_\times$).  By symmetry, the values $R^\prime_{{0}\uparrow},R^\prime_{{0}\downarrow} $ lie a distance
\begin{equation}
|\delta R_E - \delta R_{QQ}/2| = \Rxzero \pm \RxzeroUncert \mathrm{\,ppm}
\end{equation}
 either side of the crossing point, i.e.\ at  $R^\prime = $ \RzeroUp, \RzeroDown\,ppm respectively.   They therefore share a common EDM value $d_0$.  Using our fitted function from \secabbr \ref{sec:grav_depol}, we find that this shift in $R^\prime$ away from $R_\times$ requires us to add $\delta d = (\dnfQEcorrection \pm \dnfQEUncert) \times10^{-26}$ \ecm\ to our earlier crossing-point value $d_\times = (\dxValue \pm \dxUncertChisqCorr) \times10^{-26}$ \ecm\  (where we have expanded the uncertainty on $d_\times$ by $\sqrt{\chi_\nu^2}$ as we did for $R_\times$), yielding   
$d_0 = (\dzero \pm \dzeroUncert) \times 10^{-26}$ \ecm\ as the EDM value corrected for quadrupole shifts and Earth's rotation.

Further corrections must now be made to accommodate the dipole field.  The formulation of Pignol and Roccia \cite{Pignol_2012} predicts that the dipole characterized by the parameters of our fit would produce, in our storage bottle with its door cavity, a false EDM of $d_\mathrm{dip} = \ddip \times10^{-26}$ \ecm.  The crossing lines of \figabbr \ref{fig:edm_vs_R} are therefore higher by this amount than they would otherwise be, and we must compensate by subtracting $d_\mathrm{dip}$ from all of the measured data, and thus, ultimately, from $d_\times$.

In addition, this dipole field applies a volume-averaged magnetic-field gradient $\partial B_z/\partial z$ of 0.26 nT/m, which shifts the normalized frequency ratio $R^\prime$ by $\delta R_\mathrm{dg} = \pm \Rdg$\,ppm (where the subscript ``dg'' indicates ``dipole gradient'', and the upper sign once again corresponds to $\vecB$ down).  This shift of the lines in opposite directions moves the crossing point downwards by $ \ddg  \times10^{-26}$ \ecm.  The net dipole-related correction that must be applied is therefore now $(\deltaDipCorrection \pm \deltaDipUncert)  \times10^{-26}$ \ecm.  As mentioned above, the dipole strength and position arising from the fit are strongly correlated, so the uncertainty on this total dipole-related shift was calculated by studying the behaviour of $\chi^2$ as a function of the shift for a variety of dipole strengths and positions.  

We also at this point add in quadrature to the uncertainty the contribution of $0.6 \times 10^{-26}$ \ecm\ from a possible mercury-door PMD, as discussed in \secabbr \ref{sec:dip_enhance}.

Taking all of these dipole-related components into account, the net EDM thus far -- compensated for the false-EDM effects arising from field gradients, dipole and quadrupole fields, and Earth's rotation -- is    $d_\mathrm{fec} = (\dfec \pm \dfecUncert) \times10^{-26}$ \ecm, where the subscript ``fec'' stands for field-effect compensated. 


\subsection{Consistency check: Polarization data}




Because UCNs of different energies have different values of $\Delta h$, the surviving UCN polarization within a trap decreases in proportion to $(\dBzdz)^2$.  One would therefore expect that $\alpha$ would be maximized close to the $R^\prime$ value (for each $\vecB$ direction) at which $\langle\dBzdz\rangle$ is zero, as given by \eqnabbr \ref{eqn:Rprime0} with an additional correction for the dipole shift $\delta R_\mathrm{dg}$.  In fact, our calculations show that, due to the dipole field, $(\dBzdz)^2$ is minimized at \AlphaPeakShiftFromR0\,ppm below $R^\prime_{0\uparrow}$ for $\vecB$ up, and \AlphaPeakShiftFromR0\,ppm above $R^\prime_{0\downarrow}$ for $\vecB$ down.    We therefore expect the polarization $\alpha$ to be maximized at $R^\prime = \AlphaPeakUp$\,ppm, $\AlphaPeakDown$\,ppm for $\vecB$ up, down respectively.

\begin{figure}[ht]
  \begin{center}
   \resizebox*{0.5\textwidth}{!}{\includegraphics  {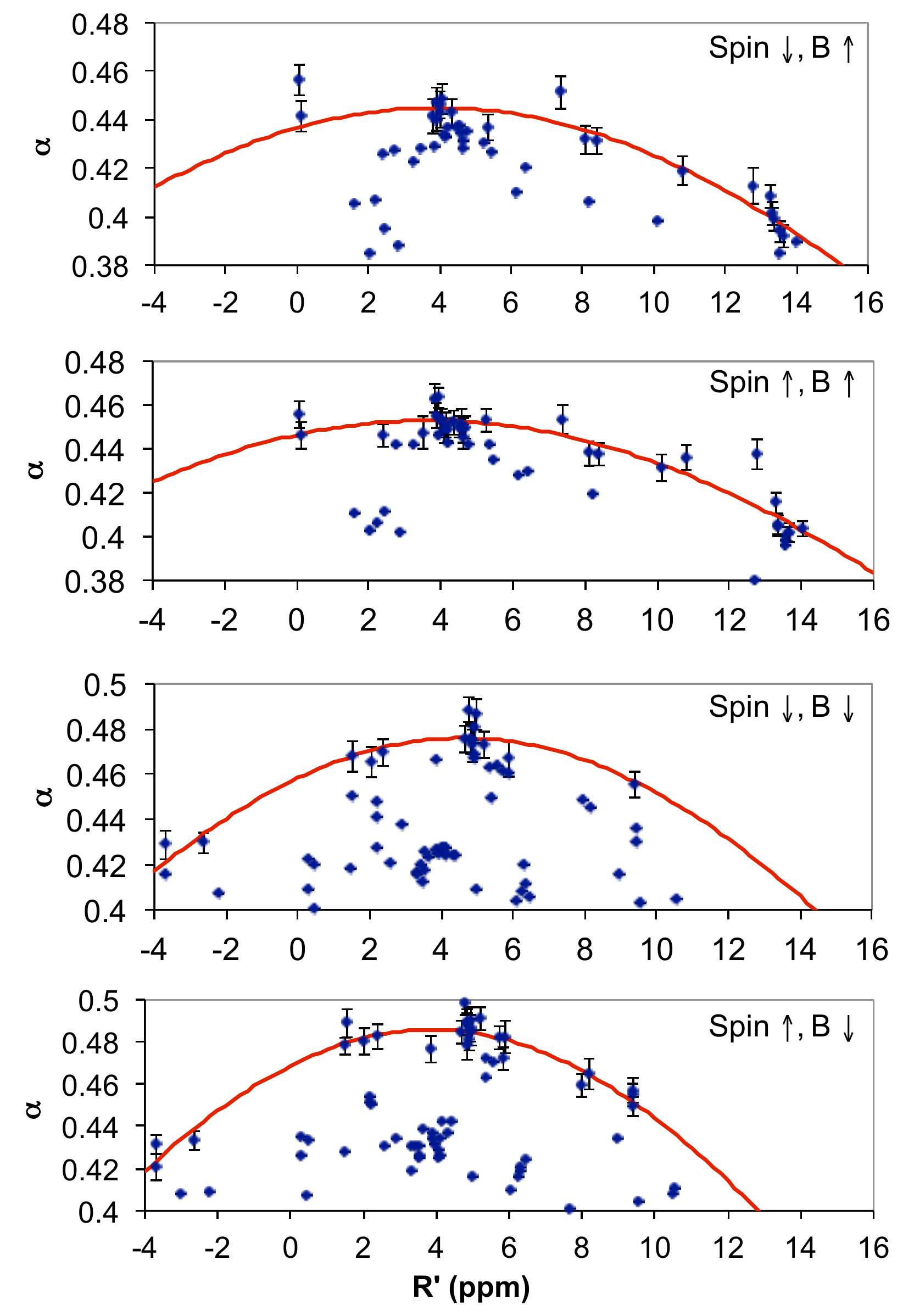}}
    \caption{(Color online)  Polarization product $\alpha$ as a function of $R^\prime$, for (a) $\vecB$ up, spin down; (b) $\vecB$ up, spin up; (c) $\vecB$ down, spin down, and (d) $\vecB$ down, spin up. Solid lines are Gaussian fits to points along the upper envelope of the data set (which are shown with error bars).}
    \label{fig:alpha_peaks}
  \end{center}
\end{figure}

During runs prior to January 2000 the gradient was varied frequently and over a fairly wide range.  We have plotted in \figabbr \ref{fig:alpha_peaks} the polarization $\alpha$ as a function of $R^\prime$ for this subset of runs, which represent 14\% of the total statistical weight of the EDM data analyzed here. Although other factors (e.g.\ loss of polarization during filling) affect the data to a greater or lesser extent, the envelope of these data points ought to peak in the region where the magnetic-field gradient is minimized.

ln the 2006 analysis, bands of points along the tops of these distributions were fitted to Gaussians to find the positions of the peaks.  
The fit was carried out multiple times, each time repeatedly removing points that lay more than $p\sigma$ below the curve, where $p$ was varied between 4.0 and 1.0, such that the final remaining points would represent the envelope. In recognition of other factors contributing to depolarization, and in order not to allow the few data points with particularly small uncertainties on $\alpha$ to skew the fits unduly, an additional uncertainty of 1\% was added in quadrature to each error bar. Points with error bars are those used in the fits represented by the solid lines.   

Although these results were used in the global fit in the 2006 analysis, we feel upon reflection that the large scatter of data points underneath each peak casts doubt upon their validity for this purpose.   In addition, the fitted peak position has some dependence both upon the number of data points included and upon the functional form of the fitted curve -- which, as we now understand \cite{Afach_2015d}, can resemble a sharp peak rather than a Gaussian.  We therefore present them here only as a consistency check, and note that, although the $R^\prime$ values at which the fitted curves peak appear to be arguably just a little higher than expected, they are close enough to be regarded as being in broad agreement.

\section{Other systematic errors}
\label{sec:other_syst}

We now consider systematic errors that do not involve the field-gradient induced false-EDM effect, and that do not bias the results.  For those that are discussed in detail in \cite{Baker_2014}, we summarise the nature of the effect and state their contributions to the overall uncertainty.  We begin with first-order ${\bf v}\times{\bf E}$ effects. 

\subsection{${\bf v}\times{\bf E}$ effects}

\label{sec: v cross E}

The Lorentz transformation of electric and magnetic fields to a moving
reference frame is \cite{jackson_electrodynamics} 
\begin{equation}
\vec{B}^{\prime }=\gamma \left( {\vec{B}-\frac{\vec{v}\times \vec{E}}{c^{2}}}%
\right) -\frac{\gamma ^{2}}{c^{2}(\gamma +1)}\vec{v}(\vec{v}\cdot \vec{B}),
\end{equation}
where $\gamma =1/\sqrt{1-v^{2}/c^{2}}$. As noted above (\eqnabbr \ref{eqn:vxE}), in the context of the false-EDM effect, a particle moving fairly slowly ($\gamma \approx 1$) through an electric field $\vec{E}$ therefore experiences an
additional magnetic field 
\begin{equation}
\vec{B}_v\approx {-}\frac{\vec{v}\times \vec{E}}{c^{2}} 
\end{equation}
above and beyond the laboratory magnetic field $\vec{B}_{0}.$  

 Such fields
are clearly linked to the electric field, and the necessity of controlling
the consequent spurious effects provides a strong constraint on the design
of the experiments. In particular, to generate a systematic error in the EDM
requires that $\vec{v}$, $\vec{E}$ and $\vec{B}_{0}$ have components that
are mutually perpendicular.  It was therefore necessary in this experiment to keep $\vec{E}$ and $\vec{B}_{0}$ closely aligned and the velocity $\vec{v}$ averaged over the storage time as small as possible.  Only the motion of the center of mass of the UCN gas and any net rotation of the gas about the center of mass can contribute to this first-order $\vec{v}\times \vec{E}$ effect. The effect has been clearly seen in atomic beam experiments \cite{Sandars_1964,Angel_1967,Stein_1967} and has been a cause for concern in neutron beam experiments.

If the neutrons have an average center-of-mass velocity of $\eta $ perpendicular to $\vec{E}$, and $\vec{E}$ has a component $\epsilon E$ perpendicular to $\vec{B}_{0}$, the $\vec{v}\times \vec{E}$ component of the magnetic field will give a systematic error in the EDM of 
\begin{eqnarray}
|d| &=&\frac{\mu _{n}}{c^{2}}\eta \epsilon \\
&=&1.3\eta \epsilon \times 10^{-24}~e\mathrm{\,cm}.
\end{eqnarray}

The only source of indefinitely sustained motion that can give rise to a finite average velocity would be a net upwards movement of the center of mass due to warming of the neutron gas as a result of inelastic collisions with the walls. The effects of such collisions, in which the neutron energy changes by only a small amount so that it remains in the UCN range, have been looked for by Richardson \cite{richardson_phd}. For a greased surface, an upper limit of 50~feV for the energy change per wall collision was established, from which it can be calculated that the maximum movement will be no more than 1~mm during the 130\,s Ramsey measurement period. If the volume-averaged angles between ${\bf E}$, ${\bf B}$ and ${\bf v}$ are each as high as 2$^\circ$ = 0.035 radians, the induced false EDM will be $1\times10^{-30}$ \ecm.  (Note that such warming is distinct from the changing $\Delta h$ arising from the preferential loss of faster UCN: the latter do not ultimately contribute to the EDM measurement, and so cannot generate a systematic error.)

If there is a difference in a transverse coordinate $x$ between where the
neutrons enter the trap and their average center of mass during the
storage time, the net average velocity is $\eta =x/T_{s}$. The effect of any
such motion will be greatly reduced by the fact that it is only motion
during the storage time between the two oscillating fields that affects the
precession frequency.  Since the trap is filled for 20~s, which in our apparatus is roughly 1.3
filling-time constants, and this is followed by a 5~s settling period, the
stored neutrons have spent an average of one filling-time constant in the
trap before the first oscillating field is applied. Collisions with the
wall are expected to destroy any non-random motion within 100 bounces, i.e.,
about 10~s, so that any such motion that results from the filling process
will mostly have died away before the first oscillating field is applied.
The position of the center of mass during storage is expected to be very
close to the axis of the 67 mm diameter guide tube. In the unlikely event that the neutrons fill preferentially at $x =1$ cm towards one side of the guide tube, and ignoring any damping of the motion in the trap during
filling, we would be left with a net motion during the 130 s Ramsey measurement of up to $\sim$ 5 mm.  This effect would then contribute $2\epsilon \times 10^{-28}\ e\mathrm{\,cm} = 6\times10^{-30}$ \ecm. This dominates the earlier result associated with warming, and therefore becomes our total uncertainty for first-order  translational  $\vxE$; to be conservative we round the value up to $1\times10^{-29}$ \ecm.

Higher-order $\vxE$ effects, in particular from the net increase in $\left|{\bf B}_0\right|$ as the $\vxE$ component is added to it in quadrature \cite{Lamoreaux_1996}, contribute $< 10^{-30}$ \ecm.

In a similar manner to the translational effect, any net rotational flow of the UCN in conjunction with a radial component of the ${\bf E}$ field may lead to an induced EDM signal.  However, any such flow of UCN is expected to be attenuated by wall collisions before the first Ramsey pulse is applied. This systematic effect requires the  electric field to have a radial
component $\epsilon E_{0}$, such as may exist near the outside of the trap if the insulator of the trap becomes charged. If a net fraction $f_\mathrm{rot}$ of the neutrons have such a flow with a velocity of 1 m/s, and if the flow persists for 10~s after the storage period has begun, and hence 5~s into the Ramsey measurement, the neutrons will have moved a net $\sim$ 5 m through this field within the Ramsey measurement.   The effect on the EDM is then $5f_\mathrm{rot}\epsilon \times 10^{-24}$ \ecm. It is reasonable to assume that only a very small fraction such as $f_\mathrm{rot}\le 0.001$ could persist in orbits through this anomalous field region, and the radial field fraction $\epsilon$ is unlikely to be larger than 10\% of the primary field, giving a net error from this source of below $5\times 10^{-28}$ \ecm.  

There will also be additional cancellations to the $\vxE$ effect from the reversal of $\vecB$, but in order to be conservative we do not consider these.

\subsection{Uncompensated magnetic field fluctuations}

There may in principle be residual effects from ${\bf B}$ field fluctuations, such as hysterisis in the $\mu$-metal shield following disturbances in the stabilised 
$\vecB$ coil current supply caused by pickup from the high-voltage changes.  As discussed in \cite{Baker_2014}, we would expect this to manifest itself as a dipole-like field $B_d$ originating at the HV feedthrough.  Because of the effective height difference $\Delta h$ between the centers of the neutron and mercury systems, the experiment retains a slight vulnerability to such a systematic.  The shielding factor is expected to be about 
\begin{equation}
\frac{B_d}{\delta B_d} = \frac{r}{3\Delta h},
\end{equation}
where $r \approx $ 55\,cm is the distance from the source of the field to the center of the trap.  In our 2006 analysis, and in \cite{Baker_2014}, we used $\Delta h$ = 2.8 mm, and thereby concluded that the protection factor was about 66.  In light of our revised calculations of the spectrum, which yield $\Delta h$  =  \deltahAfterStorage mm, we now reduce this to a factor of 50.  

\begin{figure}[ht]
  \begin{center}
    \resizebox*{0.5\textwidth}{!}{\includegraphics   {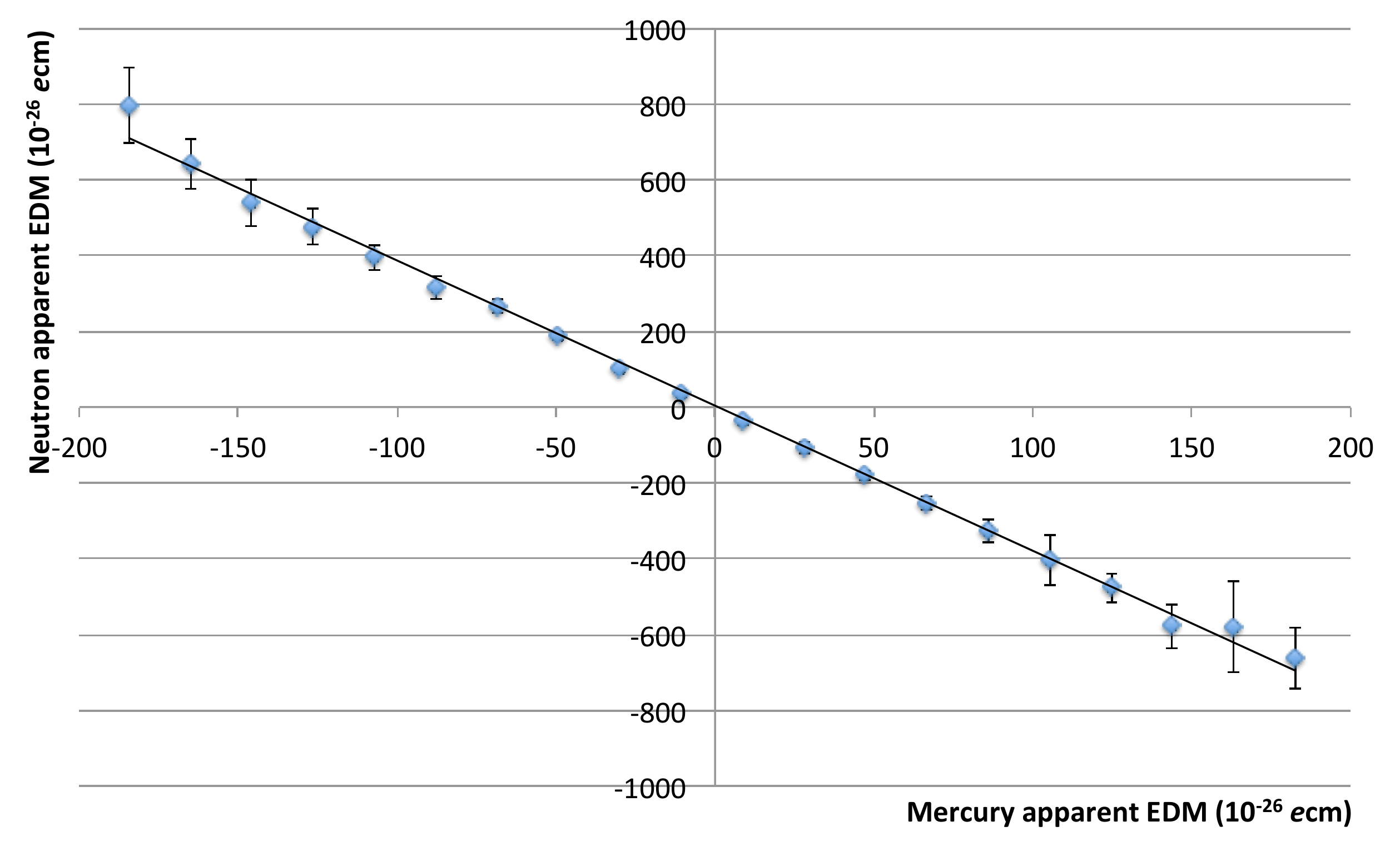}}
    \caption{(Color online) After \cite{Baker_2014} \figabbr 18. Apparent neutron EDM signals (due to uncompensated random magnetic-field fluctuations) as a function of the corresponding apparent mercury EDM signals.  Data are binned for clarity.
}
    \label{fig:n_vs_Hg_edms}
  \end{center}
\end{figure}

\figabbr 18 of \cite{Baker_2014} showed the apparent EDM signals of the neutron and mercury channels individually, plotted against one another.  \figabbr \ref{fig:n_vs_Hg_edms} here shows the same data, but this time binned for clarity.  

The neutrons alone yielded a net uncompensated EDM signal of ($17 \pm 4) \times 10^{-26}$ \ecm.  We therefore expect the mercury-magnetometer compensation to shield us from this systematic effect to a level of $(17  \times 10^{-26})/50 = 3.4  \times 10^{-27}$ \ecm.

\subsection{Mercury atom EDM}

The measured EDM values $d_\mathrm{meas}$ are obtained from a linear fit to the ratio $\nu_\mathrm{n}/\nu_\mathrm{Hg}$ versus $E$.  In principle therefore, $d_\mathrm{meas}$ contains a contribution from the intrinsic EDM $d_\mathrm{Hg}$ of the $^{199}$Hg atom.  The true $d_\mathrm{Hg}$ has been shown to be $(0.49 \pm 1.29_\mathrm{stat} \pm 0.76_\mathrm{syst}) \times 10^{-29}$ \ecm\  \cite{Griffith_2009}, so the systematic error thereby introduced into $d_\mathrm{meas}$ is a negligibly small $(-2 \pm 6) \times 10^{-29}$ \ecm.

\subsection{Electric forces}

Another possible source of systematic error arises from electrostatic forces, which may move the electrodes slightly.  Discussion of this issue in \cite{Baker_2014} concluded that the associated systematic uncertainty is  $0.4\times10^{-27}$ \ecm.

\subsection{Leakage currents}
\label{sec:syst_leakage}
\begin{figure}[ht]
  \begin{center}
    \resizebox*{0.5\textwidth}{!}{\includegraphics   {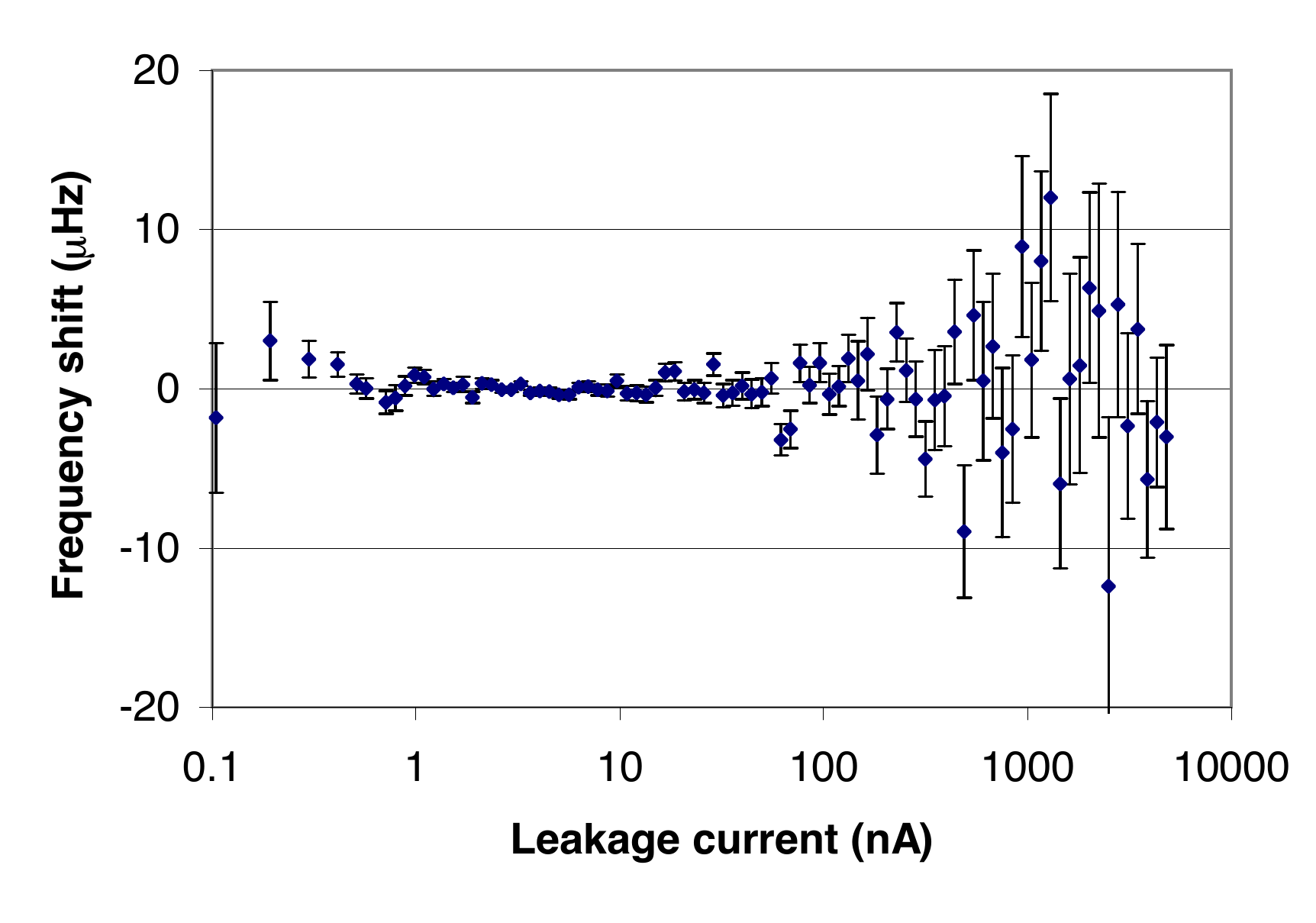}}
    \caption{(Color online) From \cite{Baker_2014}. Frequency shifts (multiplied by the polarities of the electric and magnetic fields) as a function of leakage current.}
    \label{fig:frq_shift_vs_I}
  \end{center}
\end{figure}

If the leakage current that flows through (or along the surface of)
the insulator between the electrodes has an azimuthal component, a component of the magnetic
field due to the current would be parallel (or anti-parallel) to $\vec{B}_{0}$ and
would produce a frequency shift that changes sign when the polarity of the
electric field is reversed, giving rise to a systematic error in the EDM.  As discussed in \cite{Baker_2014}, 
the false signal that would result is likely to be no larger than $0.1\times 10^{-27}$ \ecm.  As is clear from \figabbr \ref{fig:frq_shift_vs_I}, which is drawn from \cite{Baker_2014}, no dependence of frequency shift upon the leakage current is apparent in the data.

\subsection{Sparks}

High-voltage breakdown within the apparatus can produce localised high
current densities, and can in principle lead to permanent changes in the
residual magnetization of the magnetic shield. If these sparks occur
preferentially for one direction of the electric field, systematic effects
could be induced in the precession frequency, and hence in the EDM signal. A
similar effect could be produced by hysteresis in the innermost shield
following a disturbance to the $B_{0}$ supply. However, the mercury would
naturally compensate for any such effect, just as with any other shifts in
the magnetic field.

As sparks invariably disrupt the mercury frequency measurement, cycles that
contain them are excluded from the analysis, so beyond the residual effects
just discussed the sparks themselves cannot contribute to any artificial EDM
signals.  We therefore do not associate any additional systematic uncertainty to this effect.

\subsection{HV AC ripple}

A ``ripple'' on the high voltage would generate an oscillating displacement current in the storage chamber and thereby an  oscillating $B$ field. Through the Ramsey-Bloch-Siegert mechanism  \cite{Ramsey_1955}, this could in principle lead to changes in precession frequency.  A detailed discussion of this effect in \cite{Baker_2014} concludes that such a mechanism would in this case not contribute to the overall uncertainty at a level higher than  $1\times10^{-29}$ \ecm.

\subsection{Effect of nonuniform mercury depolarization}

The detremental effect of the high voltage upon the mercury depolarization time could result in a false signal if (a) the average depolarization time were different for the two HV polarities, and (b) the mercury frequency had some small dependence upon the depolarization time. As discussed in \cite{Baker_2014}, evidence for such an effect was sought in the data, and it was ruled out at the level of  $1.2\times 10^{-29}$ \ecm.

\subsection{Artifacts of the measurement process}

In this class of systematics there is usually no real change in precession frequency ---
only an apparent change due to a malfunction of the measurement process. An
example might be bursts of false counts in the neutron detection channel
derived from a high-voltage spark. No evidence of any such effects has been
seen, and in any case any artificial signals from such effects would tend to cancel upon reversal of $\vecB$.  We do not associate any further uncertainty with this mechanism.

\subsection{Accuracy of Hg frequency measurements}
\label{sec:Hg_accuracy}

A number of mechanisms can affect the frequency measurement of the Hg magnetometer.  These are discussed extensively in \cite{Baker_2014}, and are already implicitly dealt with by our analysis.  No further uncertainty is assigned to any of  these effects.

\subsection{Stability of results}

\begin{figure} []
\begin{center}
\resizebox{0.5\textwidth}{!}{
\includegraphics{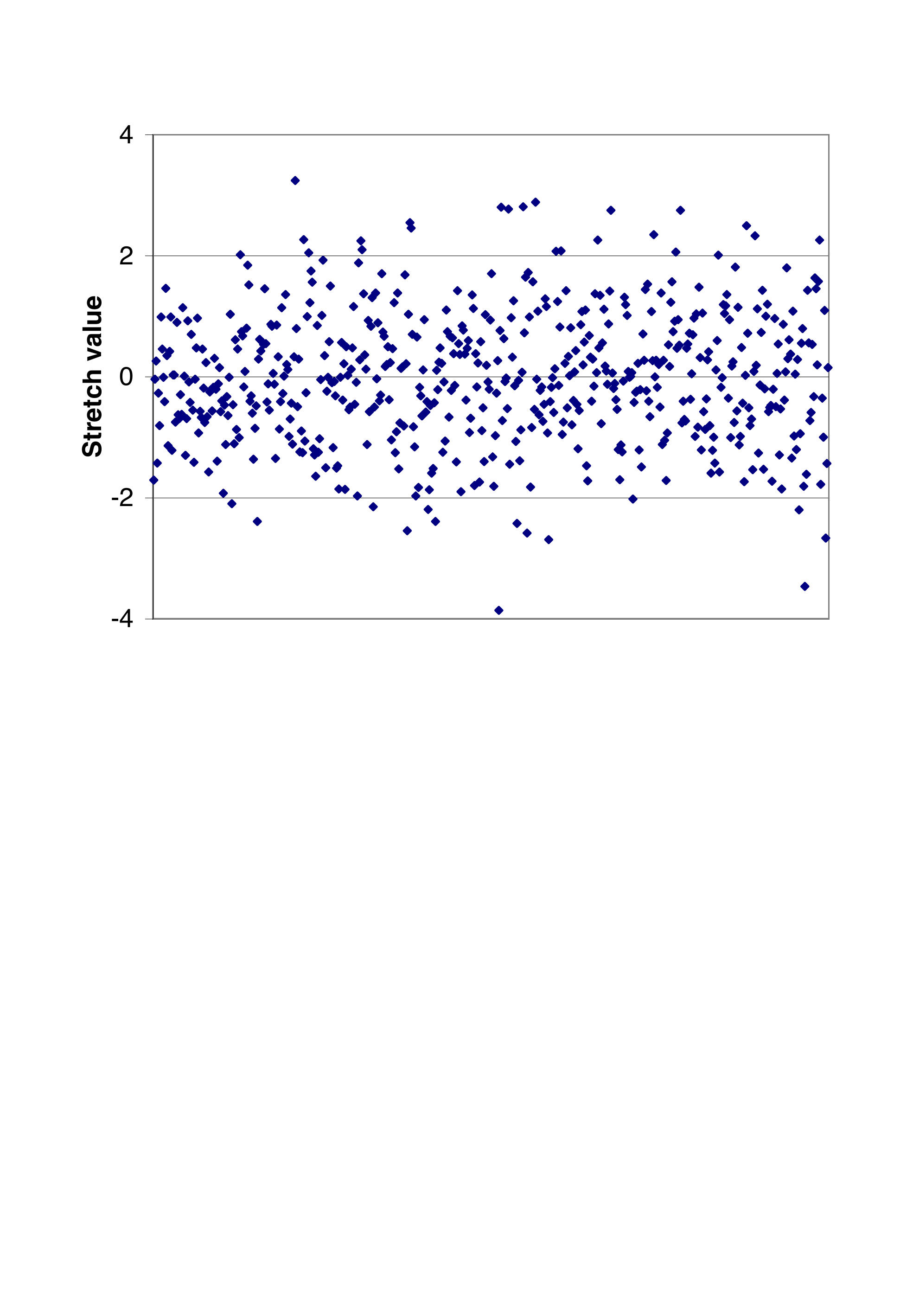}}
\end{center}
\caption{(Color online) Stretch values from the fits of \figabbr \ref{fig:edm_vs_R} in the order in which the data were taken}
\label{fig:stability}
\end{figure}

As a final check for possible unforeseen systematic errors, we consider the stability of the measurements.  
On a run-by-run basis, one can consider the stretch values (\eqnabbr \ref{eqn:gp_stretch}) from the geometric-phase line fit of \figabbr \ref{fig:edm_vs_R} as a function of time, as shown in \figabbr \ref{fig:stability}: this removes the natural variation in the measured EDM due to the changing $R_a$ values.   If there were an offset in one series of data due, for example, to a magnetic anomaly that was later removed, one would see points typically above the line for a series of runs at one period in time and below the line at other times.  No such trend is visible.


\subsection{Systematic error summary}
   
\tababbr \tabnumSystErrs\ lists all of the systematic errors that we have discussed, and gives their uncertainties.  Its last line specifies the net shift in EDM value, and the total systematic uncertainty.  For the avoidance of doubt, we note that the shifts listed represent the offsets generated by the biases in question.  The compensating corrections that have been applied are therefore equal in magnitude and opposite in sign to the stated values.

\begin{table}
\label{tab:syst_summary}
	\begin{center}
		\begin{tabular}{|l|l|l|}
		\hline
		{\bf Effect}  & {\bf Shift}& {\bf $\sigma$}\\
		\hline 
		$\nu_\mathrm{Hg}$ light shift (included in $d_\times$)& (0.35) & 0.08\\
		\hline
		$\chi_\nu^2$ = 1.2 adjustment 			& 0 		&\dxChisqAdjustment \\
		\hline
		Quadrupole fields \& Earth rot'n            &  \dnfQEshift	& \dnfQEUncert     \\
		\hline
		Dipole field					& \deltaDipShift		& \deltaDipUncert	\\
		\hline
		Hg door PMD                        		& 0.00 	& 0.60 \\
		\hline
		${\bf v}\times{\bf E}$ translational 	& 0.000 & 0.001\\
		\hline
		${\bf v}\times{\bf E}$ rotational          	& 0.00 & 0.05\\
		\hline
		Second-order ${\bf v}\times{\bf E}$       & 0.000 & 0.000\\
		\hline
		Uncompensated B drift                      	& 0.00 & 0.34\\
		\hline
		Hg atom EDM                                	& -0.002 & 0.006\\
		\hline
		Electric forces                            		& 0.00 & 0.04\\
		\hline
		Leakage currents                           	& 0.00 & 0.01\\
		\hline
		AC fields                                  		& 0.000 & 0.001\\
		\hline
		Nonuniform Hg depol.                           & 0.000 & 0.001\\
		\hline
		{\bf Total  shift of $\dx$}                                & {\bf \totalShiftExceptLight} & {\bf \totalShiftUncert}\\
		\hline
		\end{tabular}
	\end{center}
\caption{Summary of systematic errors and their uncertainties, in units of 10$^{-26} ~e$cm.  Correction for the mercury light shift is already incorporated run by run prior to the crossing-lines fit; other corrections are then applied to the crossing-point EDM value $d_\times$. }
\end{table}

\section{Results and conclusion}
\label{sec:final_result}

We have reanalyzed the data that were used in 2006 to calculate an upper limit on the magnitude of the neutron electric dipole moment \cite{Baker_2006}.  As in the 2006 analysis, earlier data from  the interim result of \cite{Harris_1999} have been excluded.    We have here taken into account recent developments, including an understanding of the process of gravitational depolarization and more detailed calculations of the spectrum of the stored UCN, and we have also taken into consideration measurements made in the meantime on the apparatus. The picture that emerges appears to be perfectly self-consistent.

Once the crossing-point EDM value $d_\times = (\dxValue \pm \dxUncert) \times 10^{-26}$ \ecm\ is corrected for the systematic biases listed in \tababbr \tabnumSystErrs, as shown explicitly in \tababbr III for each stage of the analysis, we obtain (by adding in quadrature the statistical and systematic uncertainties) a final value of $d_\mathrm{n} = (\finalEdm \pm \finalUncert) \times10^{-26}$ \ecm.  The 90\% and 95\% confidence limit (CL) ranges are therefore $\rangeNinetyLower < d_\mathrm{n} < \rangeNinetyUpper \times10^{-26}$ \ecm\ and  $\rangeNinetyfiveLower < d_\mathrm{n} < \rangeNinetyfiveUpper \times10^{-26}$ \ecm\ respectively.  Taking limits that are symmetric about zero yields $\left| d_\mathrm{n} \right| < \limitNinety  \times10^{-26}$ \ecm\ (90\% CL) and $\left| d_\mathrm{n} \right| < \limitNinetyfive  \times10^{-26}$ \ecm\ (95\% CL).

\begin{table}
\label{tab:d_stages}
	\begin{center}
		\begin{tabular}{|l|l|l|l|l|}
		\hline
		{\bf Analysis stage}  &{\bf EDM} &  {\bf $\sigma$} \\
		\hline 
		Crossing point $d_\times$ 			&\dxValue&\dxUncert \\
		\hline
		Gradient-corrected $d_0$            		& \dzero & \dzeroUncert  \\
		\hline
		Dipole-corrected $d_\mathrm{fec}$      &\dfec & \dfecUncert\\
		\hline
		{\bf Final result $d_n$}                                &{\bf \dfec}&{\bf \finalUncert} \\
		\hline
		\end{tabular}
	\end{center}
\caption{Summary of the net EDM arising from each stage of the analysis, in units of 10$^{-26}$ \ecm.}
\end{table}

This analysis was also repeated with a number of variations: for example, with the spectrum derived from the softened-Maxwell approximation, which yielded (with the same uncertainty) central values of 0.86 and 0.79 $ \times10^{-26}$ \ecm\ with the slope multiplier $k^\prime$ kept fixed and allowed to vary, respectively; and without the apparently anomalous Run 1900 (resulting in central values of -0.05 $ \times10^{-26}$ \ecm\ with the simulated spectrum, and +0.43 $ \times10^{-26}$ \ecm\ with the softened-Maxwell spectrum).  Furthermore, we have also looked carefully at the implications of the fields as defined by the maps produced by the 2010 scans at PSI.  We do not consider it appropriate to apply the results directly to our data, not only because the field scans were carried out after a lapse of several years and following dismantling, shipping and reassembly of the shields, but also because the accuracy of measurements of small transverse field components, typically a fraction of a percent of the main vertical field strength, is limited by  the mechanical precision of the mapper system itself.  Nonetheless, we considered it useful to carry out the analysis with the quadrupole shifts $\delta R_Q$ fixed at the values defined by these more recent field measurements.   The resulting value of $d_n$ was  $0.76 \times10^{-26}$ \ecm\ (again with the same uncertainty), although as might be expected $\chi^2/\nu$ for the global fit deteriorated somewhat to 1.48.  The range of variation of all of these results is of the order of the uncertainty on our systematic errors, and all lie comfortably within our confidence-limit range.

 Ultimately, and in conclusion, after a thorough re-evaluation of the previously known systematic effects and the inclusion of various other newly established effects,  the overall corrections to the 2006 result have been found -- both individually and collectively -- to be relatively small compared with the statistical uncertainty. As a direct consequence, our new limit represents a departure of only a few percent from the earlier limit of $\left| d_\mathrm{n} \right| < 2.9  \times10^{-26}$ \ecm\ (90\% CL) \cite{Baker_2006} despite the very real differences between the two analyses.



At the time of writing, the apparatus employed in this experiment is enjoying something of a renaissance.  It has been very substantially upgraded, including in particular with the provision of an array of high-precision cesium magnetometers  \cite{Knowles_2009}, for a further nEDM measurement \cite{Baker_2011} at the new UCN source at PSI \cite{Lauss_2014}.  Furthermore, an innovative new spin-echo based technique \cite{afach15b} allows both the measurement of the UCN spectrum within the apparatus as well as an accurate determination of magnetic-field gradients.  It is currently running with a greater sensitivity than ever before, and, with potential systematic errors now being very tightly constrained, new results will be limited only by statistics for some time to come.

\acknowledgments

This
work was funded in part by the UK Science and Technology Facilities Council (STFC) under grants ST/K001329/1, ST/M003426/1 and ST/L006472/1; the Fund for Scientific Research, Flanders; grant GO A/2010/10 of KU~Leuven; and the Swiss National Science Foundation Projects 200020-144473 (PSI), 200021-126562 (PSI), 200020-149211 (ETH) and 200020-140421 (Fribourg).  The LPC Caen and the LPSC acknowledge the support of the French Agence Nationale de la Recherche (ANR) under reference ANR-09-BLAN-0046. Polish partners wish to acknowledge support from the PL-Grid Infrastructure, Poland, and from the National Science Centre, Poland, under grant no.\ UMO-2012/04/M/ST2/00556.  The original apparatus was funded by grants from the UK's PPARC (now STFC), and we would like to thank the generations of engineers and Research Fellows who contributed to its development, including I. Kilvington, S. Lamoreaux, B. Heckel, J. Morse, Y. Chibane, and M. Chouder.  We remember with gratitute the contributions of Professor K. Smith, who led the Sussex group throughout the early years of development of the apparatus.  One of us (DJM) benefitted from an ILL studentship, and another (EW) is a PhD Fellow of the Research Foundation - Flanders (FWO).  The experiments underlying this work would not have been possible at all without the cooperation and provision of neutron facilities by the ILL itself.  We would also like to thank the PSI staff for their outstanding support during the PSI-related measurements.

Finally, the remaining authors wish to express their sorrow at the loss of Professor Mike Pendlebury, and their deep gratitude for his wide-ranging contributions to nEDM measurements worldwide.  He was an outstanding leader in the field for more than four decades.  His last message to his colleagues, just days before he passed away on 1 September 2015, was to say how pleased he was that this paper was complete.  He will be sorely missed.

\bibliography{../../2015_neutron_EDM}

\end{document}